\input epsf
\epsfverbosetrue
\documentstyle{mn}

\newif\ifAMStwofonts
\ifoldfss
  \ifCUPmtlplainloaded \else
    \NewTextAlphabet{textbfit} {cmbxti10} {}
    \NewTextAlphabet{textbfss} {cmssbx10} {}
    \NewMathAlphabet{mathbfit} {cmbxti10} {} % for math mode
    \NewMathAlphabet{mathbfss} {cmssbx10} {} %  "   "    "
  \fi
  \ifAMStwofonts
    \ifCUPmtlplainloaded \else
      \NewSymbolFont{upmath} {eurm10}
      \NewSymbolFont{AMSa} {msam10}
      \NewMathSymbol{\upi}     {0}{upmath}{19}
      \NewMathSymbol{\umu}     {0}{upmath}{16}
      \NewMathSymbol{\upartial}{0}{upmath}{40}
      \NewMathSymbol{\leqslant}{3}{AMSa}{36}
      \NewMathSymbol{\geqslant}{3}{AMSa}{3E}

      \let\leq=\leqslant 
      \let\geq=\geqslant 
    \fi
  \fi
\fi % End of OFSS

\ifnfssone
  \newmathalphabet{\mathit}
  \addtoversion{normal}{\mathit}{cmr}{m}{it}
  \addtoversion{bold}{\mathit}{cmr}{bx}{it}
  \newmathalphabet{\mathbfit} % math mode version of \textbfit{..}
  \addtoversion{normal}{\mathbfit}{cmr}{bx}{it}
  \addtoversion{bold}{\mathbfit}{cmr}{bx}{it}
  \newmathalphabet{\mathbfss} % math mode version of \textbfss{..}
  \addtoversion{normal}{\mathbfss}{cmss}{bx}{n}
  \addtoversion{bold}{\mathbfss}{cmss}{bx}{n}
  \ifAMStwofonts
    \ifCUPmtlplainloaded \else
      \UseAMStwoboldmath
      \makeatletter
      \new@mathgroup\upmath@group
      \define@mathgroup\mv@normal\upmath@group{eur}{m}{n}
      \define@mathgroup\mv@bold\upmath@group{eur}{b}{n}
      \edef\UPM{\hexnumber\upmath@group}
      \new@mathgroup\amsa@group
      \define@mathgroup\mv@normal\amsa@group{msa}{m}{n}
      \define@mathgroup\mv@bold\amsa@group{msa}{m}{n}
      \edef\AMSa{\hexnumber\amsa@group}
      \makeatother
      \mathchardef\upi="0\UPM19
      \mathchardef\umu="0\UPM16
      \mathchardef\upartial="0\UPM40
      \mathchardef\leqslant="3\AMSa36
      \mathchardef\geqslant="3\AMSa3E

      \let\leq=\leqslant 
      \let\geq=\geqslant 
    \fi
  \fi
\fi % End of NFSS release 1

\ifnfsstwo
  \DeclareMathAlphabet{\mathbfit}{OT1}{cmr}{bx}{it}
  \SetMathAlphabet\mathbfit{bold}{OT1}{cmr}{bx}{it}
  \DeclareMathAlphabet{\mathbfss}{OT1}{cmss}{bx}{n}
  \SetMathAlphabet\mathbfss{bold}{OT1}{cmss}{bx}{n}
  \ifAMStwofonts
    \ifCUPmtlplainloaded \else
      \DeclareSymbolFont{UPM}{U}{eur}{m}{n}
      \SetSymbolFont{UPM}{bold}{U}{eur}{b}{n}
      \DeclareSymbolFont{AMSa}{U}{msa}{m}{n}
      \DeclareMathSymbol{\upi}{0}{UPM}{"19}
      \DeclareMathSymbol{\umu}{0}{UPM}{"16}
      \DeclareMathSymbol{\upartial}{0}{UPM}{"40}
      \DeclareMathSymbol{\leqslant}{3}{AMSa}{"36}
      \DeclareMathSymbol{\geqslant}{3}{AMSa}{"3E}

      \let\leq=\leqslant 
      \let\geq=\geqslant 
    \fi
  \fi
\fi % End of NFSS release 2

\ifCUPmtlplainloaded \else
  \ifAMStwofonts \else % If no AMS fonts
    \def\upi{\pi}
    \def\umu{\mu}
    \def\upartial{\partial}
  \fi
\fi

\title{Supernovae in the nuclear regions of starburst galaxies}

\author[S.Mattila and W.P.S.Meikle]
       {S. Mattila\thanks{Send offprint requests to: s.mattila@ic.ac.uk} and W.P.S.Meikle \\ 
Astrophysics Group, Blackett Laboratory, Imperial College of
Science, Technology and Medicine, Prince Consort Road, \\ London 
SW7 2BW \\
}
\date{Accepted 2000 November 22.
      Received 2000 October 20;
      in original form 2000 May 24}  
\pagerange{\pageref{firstpage}--\pageref{lastpage}}
\pubyear{1994}

\begin{document}

\maketitle

\label{firstpage}

\begin{abstract}
The feasibility of using near-infrared observations to discover
supernovae in the nuclear and circumnuclear regions of nearby
starburst galaxies is investigated.  We provide updated estimates of
the intrinsic core-collapse supernova rates in these regions.  We discuss the 
problem of extinction, and present new estimates of the extinction towards 33
supernova remnants in the starburst galaxy M~82.  This is done using
H~I and H$_{2}$ column density measurements.  We estimate the molecular to 
atomic hydrogen mass ratio to be 7.4 $\pm$ 1.0 in M~82.  We have assembled 
near-infrared photometric data for a total of 13 core-collapse supernovae, some
unpublished hitherto.  This constitutes the largest database of IR
light curves for such events.  We show that the IR light curves fall
into two classes, ``ordinary'' and ``slow-declining''.  Template
$JHKL$ light curves are derived for both classes.  For ordinary
core-collapse supernovae, the average peak $JHKL$ absolute magnitudes
are --18.4, --18.6, --18.6, and --19.0 respectively.  The
slow-declining core-collapse SNe are found to be significantly more
luminous than the ordinary events, even at early times, having average
peak $JHKL$ absolute magnitudes of --19.9, --20.0, --20.0, and --20.4
respectively.  We investigate the efficiency of a computerised image
subtraction method in supernova detection.  We then carry out a Monte
Carlo simulation of a supernova search using $K$-band images of
NGC~5962.  The effects of extinction and observing strategy are discussed.  We 
conclude that a modest observational programme will be able to discover a number of 
nuclear supernovae.
\end{abstract}

\begin{keywords}
supernovae -- starburst galaxies -- image subtraction.
\end{keywords}

\section{Introduction}
Core-collapse supernovae (SNe) are observed to occur in sites of
recent star formation.  Such regions contain large quantities of dust,
especially the nuclear ($r$ $\la$ 500~pc) regions of
starburst galaxies.  Consequently, SN search programmes working at
optical wavelengths probably miss a significant number of events in
starburst galaxies due to dust obscuration.  Thus, the observed
supernova rates may be only a lower limit for the true supernova rate.
The inferred star formation rates and radio observations of young
supernova remnants (SNRs) indicate that highly-extinguished supernovae
should exist in the nuclear regions of nearby starburst galaxies.  In
NGC~253 ({\it e.g.} Ulvestad $\&$ Antonucci 1997), NGC~2146 (Tarchi
{\it et al.}  2000), M~82 ({\it e.g.} Allen $\&$ Kronberg 1998) and
NGC4038/39 (Neff \& Ulvestad 2000), several compact non-thermal radio
sources have been identified as young SNRs indicating explosion rates
of obscured supernovae of around 0.1~yr$^{-1}$.  Also, high resolution
radio observations have recently revealed a group of luminous radio
supernovae in the nuclear regions of ARP~220 (Smith {\it et al.}
1998). Yet only one supernova event (SN 1940E in NGC 253) has ever been 
directly observed in (or in front of) the nuclear regions in any of these 
five galaxies.

It is desirable to discover and study SNe in the centres of nearby
starburst galaxies.  In addition to providing a better estimate of the
core-collapse rate in the local universe, such a survey would offer an
opportunity to study the behaviour of supernovae in the dusty and
dense nuclear environment and to probe the nuclear extinction.
Indeed, SNe exploding in the nuclear regions of galaxies
may differ significantly from ordinary SNe due to the higher density
environment as suggested in the starburst model for AGN (Terlevich
{\it et al.}  1992).  The best-studied example of a core-collapse SN
exploding within a high density circumstellar environment is SN 1998S 
(Fassia {\it et al.} 2000a, Fassia {\it et al.} 2000b, Gerardy {\it et al.}
2000, Leonard {\it et al.} 2000).  Ultimately, such a survey would also be 
valuable in guiding attempts to determine the rates of high redshift
core-collapse supernovae.  Such a high redshift study is one of the
aims of the NGST (New Generation Space Telescope) (Jorgensen {\it et
al.} 1997; Madau {\it et al.}  1998, Dahlen \& Fransson 1999; Sullivan
{\it et al.} 2000).  However, before starting to investigate the
evolution of the SN rate with redshift it is important to have an
accurate measurement of the local SN rate and to understand the
behaviour of supernovae within the dusty, high density starburst
regions of galaxies.

The feasibility of searching for obscured supernovae in nuclear
starburst regions has been discussed in two papers.
Van Buren \& Norman (1989) considered the use of mid-IR narrow-band 
(10.52 $\mu$m) imaging.  They estimated that supernovae would be observable 
in the mid-IR with ground-based 4-10 m telescopes at distances of 
15-40 Mpc in one night's integration. Supernova detection in the $K$-band was 
studied by Grossan {\it et al.} (1999). They emphasised the need for 
high spatial resolution to allow successful image subtraction in the nuclear 
regions of galaxies.  

An optical search for supernovae in a sample of 142 nearby
star-forming galaxies was carried out in 1988-1991 by Richmond {\it et
al.} (1998).  The observations were carried out with a $500\times500$
pixel CCD camera (0.58"/pixel) on a 1~m telescope at the Lick
Observatory.  Two search procedures were used.  In the first method,
the search images were compared by eye.  This yielded five supernovae,
all of which were outside the host galaxies' extinguished nuclear
regions.  The deduced SN rates were similar to those measured in
normal galaxies.  In the second method, differential photometry was
carried out on the nuclei of the galaxies.  However, they did not
discover any brightening attributable to nuclear supernovae.  From
this, they deduced upper limits for the {\it unobscured} SN rates
within the nuclear regions {\it viz.} $<$ 9h$^{2}$~SRU\footnote{ One
SRU is the number of supernovae per century per 10$^{10}$L$_{\odot}$
galaxy blue luminosity, and H$_{\rm 0}$ = h $\times$ 100 kms$^{-1}$/Mpc.},
$<$ 12h$^{2}$~SRU, and $<$ 7h$^{2}$~SRU for Types Ia, Ib/c, and II,
respectively.  This work confirmed that, given the likely high
obscuration in nuclear starbursts, for a search for supernovae in such
regions to be successful it would have to be carried out at infrared
wavelengths.

Van Buren {\it et al.} (1994) conducted a $K$-band survey for
supernovae in starburst galaxies at NASA's Infrared Telescope
Facility, Hawaii for about 2 years. Most of the observations were made
with PROTOCam, and the Richardson-Lucy image restoration algorithm was
used to enhance the resolution of the survey images.  They discovered
SN~1992bu in NGC~3690 in $K$-band images$\footnote{
www.ipac.caltech.edu/ipac/info/sn1992/sn1992\_NGC3690.html}$.  The
supernova is located $\sim$6" from Core B1 of the galaxy. They
measured $K$-band magnitudes of +16.6, +17.2, and +18.1 in three
images separated by about a month.  No other observation of this
supernova was reported, and thus it is not possible to say if it was 
located within the obscured nuclear regions. 

A more recent (1992-1994) near-infrared (NIR) search for supernovae in
177 nearby IRAS galaxies has been carried out in the $K'$-band by
Grossan {\it et al.} (1999). The observations were performed with the
2.3~m Wyoming IR Observatory (WIRO) telescope using the Michigan IR
camera (MIRC), with a 2.2"/pixel resolution and 128$\times$128
pixels. The seeing was typically between 0.7'' and 1.0" and so the
seeing disk was usually undersampled. The length of the time between
two observations of a given galaxy varied typically from 1 to 3
months. They did not discover any supernovae during the period,
limiting the SN rate {\it outside the nuclear regions} ($>$ 15" radius) to 
less than 3.8 FIRSRU$\footnote{ One FIRSRU is the number of supernovae per 
century per 10$^{10}$L$_{\odot}$ galaxy far-IR luminosity (see Section~2).}$
in their sample galaxies within 25~Mpc distance.  They concluded that this 
negative result was due to the poor resolution of the camera and that a 
higher resolution NIR supernova search covering only the inner $\sim$450pc of 
each galaxy would be more productive.

The latest reported attempt to detect SN explosions in the nuclear
regions of starburst galaxies was by Bregman {\it et al.} (2000). They
observed a sample of 10 galaxies within 7~Mpc distance using ISOCAM
with a pixel size of 3''. After continuum subtraction they looked for
traces of [Ni~II] 6.63 $\mu$m line emission which would have indicated
recent SN explosions.  However, they did not detect any [Ni~II]
emission in the sample galaxies.  They presented an upper limit for the
SN rate in their galaxy sample normalised to M~82 viz. 0.3~yr$^{-1}$.
Given the far-IR luminosity of M~82 (Appendix~A) this corresponds to 7.2 
FIRSRU.

Although, as argued above, the discovery of such SN events would
clearly be important, given the uncertainties in {\it e.g.} nuclear
extinction or SN magnitudes it is less obvious what might be achieved
in practice with existing telescopes.  We have therefore carried out a
study to test the feasibility of discovering obscured supernovae in
the nuclear regions of starburst galaxies.  In this paper we report
the results of this study.  In Section~2, we review indirect methods
of assessing the nuclear supernova rate and provide an updated
estimate of this parameter.  In Section~3, extinction towards nuclear
supernovae is discussed.  In particular, we present a new estimate of
the extinction and its distribution in the starburst galaxy M~82.  In
Section~4, we present $JHKL$-band light curves for a total of 13
events, some unpublished hitherto.  This constitutes the most complete
database of core-collapse SN IR light curves ever published.  We show
that the IR light curves fall into two classes, ``ordinary'' and
``slow-declining''.  Template $JHKL$ light curves are derived for both
classes.  In section 5, supernova detection using an image subtraction
method is described and tested, and in section~6 Monte Carlo simulations are 
used to test the feasibility of a practical supernova search.  The results 
are summarised in Section~7.

\section{Indirect estimation of the nuclear supernova rates}

A wide variety of indirect methods have been employed to estimate the
nuclear SN rates for nearby starburst galaxies.  These
include (a) observations of young SNRs using radio interferometry, (b)
measurement of NIR [Fe~II] luminosities of nuclear regions, and (c)
measurement of non-thermal radio luminosities of nuclear regions.  A
collection of SN rate estimates from the literature is given in Table
$\ref{rates}$ for M~82, NGC~253, and NGC~4038/39. 

\begin{table*}
 \centering
 \begin{minipage}{140mm}
  \caption{Indirect estimates of starburst nuclear supernova rates}
  \begin{tabular}{@{}llll@{}} \hline
Galaxy & $r_{\rm SN}$ (yr$^{-1}$) & Method & Author \\
\hline
M82	& $>$0.016 & SNR source counts, radio source ages & Allen $\&$ Kronberg (1998) \\
M82	& 0.10 & SNR source counts, radio source ages & Van Buren $\&$ Greenhouse (1994) \\
M82	& 0.05 & SNR number vs. diameter relation & Muxlow {\it et al.} (1994) \\
M82     & 0.11$\pm$0.05 & SNR number vs. diameter relation & Huang {\it et al.} (1994) \\
M82     & $<$0.10 & SNR source counts, lack of variability & Ulvestad $\&$ Antonucci (1994) \\
M82     & $\sim$0.1 & [FeII] luminosity & Greenhouse {\it et al.} (1997) \\
M82     & 0.2 & non-thermal radio luminosity & Colina $\&$ Perez-Olea (1992) \\ \hline
N253    & 0.08 &  SNR source counts, radio source ages & Van Buren $\&$ Greenhouse (1994) \\
N253   &  $<$0.30\footnote{
for the innermost 200pc radius}
 & SN source counts, lack of variability & Ulvestad $\&$ Antonucci (1997) \\
N253   & $<$0.10\footnote{
for the circumnuclear regions with radius between 200pc and 2kpc}
 & SN source counts, lack of variability & Ulvestad (2000) \\
N253   &  0.03 & [FeII] luminosity  & Engelbracht {\it et al.} (1998) \\
N253    &  0.05 &  non-thermal radio luminosity & Colina $\&$ Perez-Olea (1992)  \\ \hline
N4038/39 & $\sim$0.2 & compact non-thermal radio sources & Neff $\&$ Ulvestad (2000) \\
\hline
\label{rates}
\vspace{-0.5cm}
\end{tabular}
\end{minipage}
\end{table*}

Before examining these rate estimates, we note that the SN rates
derived from optical search programmes are often given in terms of the
SRU which has units of number of supernovae per century per
10$^{10}$L$_{\odot}$ galaxy blue luminosity (e.g. Cappellaro {\it et
al.} (1999)).  However, for nuclear starburst regions the high
extinction means that definition of the SN rate in terms of the
observed galaxy blue luminosity is rather pointless, since we cannot
measure the intrinsic blue luminosity in the region of SN occurrence.  
Most of the optical-UV energy emitted by the massive stars is absorbed and
re-radiated by dust in the FIR.  We therefore express the SN rates in terms 
of the galaxy FIR luminosity ({\it i.e.} we use FIRSRU).

We can compare directly the estimated supernova rate with $L_{\rm
FIR}$. From Table $\ref{rates}$, for M~82, NGC~253, and NGC~4038 we adopt SN
rates of 0.1~yr$^{-1}$, 0.05~yr$^{-1}$, and 0.2~yr$^{-1}$ respectively as
representative of the range of estimates.  If we then compare these
values with the galaxies' $L_{\rm FIR}$ values (see Appendix~A), we
obtain a relation between the supernova rate and FIR luminosity:

\begin{equation}
r_{\rm SN} = 2.7 \times 10^{-12} \times L_{\rm FIR} / L_{\odot}~~yr^{-1}
\end{equation}

The origin of this relation is now discussed.  The core-collapse SN rate can be
derived from the $SFR$ [M$_{\odot}$yr$^{-1}$] (e.g. Madau {\it et al.}
1998) using :

\begin{equation}
r_{SN} = \frac{\int_{m_{l}}^{m_{u}} \phi(m)
dm}{\int_{m_{min}}^{m_{max}} m\phi(m)dm} \times SFR
\end{equation}

\noindent
where $\phi$(m) is the IMF with lower and upper mass cut-offs of
$m_{min}$ and $m_{max}$, and where $m_{l}$ and $m_{u}$ are the lower
and upper mass limits for a core-collapse SN progenitor. For SN
progenitor masses between 8 and 50~M$_{\odot}$ (Tsujimoto {\it et al.} 1997)
and a Salpeter IMF (m$^{-2.35}$) with cut-offs of 0.1 and 125~M$_{\odot}$ 
({\it i.e.} stars of all possible masses are being formed) we obtain :

\begin{equation}
r_{\rm SN} = 0.0070 \times SFR~~yr^{-1} 
\end{equation}

Thus, if we can estimate the $SFR$ in a starburst galaxy, we shall have
an additional method for assessing the supernova rate.  The $SFR$s can be
estimated for a sample of galaxies via integrated emission line or
continuum luminosities.  A popular method for estimating the $SFR$ is
via the far--IR luminosity, $L_{\rm FIR}$, of the galaxy. This requires us
to assume that the $L_{\rm FIR}$ is powered by a population of young,
massive stars rather than by a population of old stars or an AGN.  A
number of authors have derived $SFR$s in this way, the estimates for the
ratio $SFR$/$L_{\rm FIR}$ being in the range
(1.5--6.5)$\times$10$^{-10}$~M$_{\odot}$yr$^{-1}$/L$_{\odot}$ 
(e.g. Thronson $\&$ Telesco 1986, Condon 1992, Buat $\&$ Xu 1996).

More recently Rowan-Robinson {\it et al.} (1997) indicated a relation
between $SFR$ and FIR luminosity using starburst models for optical-UV
radiation assuming that a fraction $\epsilon$ of the optical-UV energy
emitted in a starburst is absorbed and re-radiated by dust in the FIR.
If stars of all possible masses are being formed according to the
Salpeter IMF, we have :

\begin{equation}
SFR = \frac{1.5}{\epsilon} \times 10^{-10} L_{\rm FIR} /
L_{\odot}~~M_{\odot}yr^{-1}
\end{equation}

\begin{table*}
 \centering
 \begin{minipage}{140mm}
  \caption{Extinction estimates for the nuclei of starburst galaxies}
  \begin{tabular}{@{}llll@{}} \hline
Galaxy & $A_{V}$(center) & the method & the author \\ \hline
M82	& 2-12	& NIR H lines~~(screen) & Satyapal {\it et al.} (1995) \\
M82	& 10 & mid-IR line ratios~~(screen) & Genzel {\it et al.} (1998) \\
M82	& 1.19($K$) & Br$\alpha$/Br$\gamma$~~(screen) & Mouri {\it et al.} (1997) \\
M82     & 27  & NIR \& mm recomb. lines(screen)  & Puxley {\it et al.} (1989) \\
NGC253	& $\sim$9 & NIR line ratios~~(screen) & Engelbracht{\it et al.} (1998) \\
NGC253  & 17-19 & NIR colours~~(mixed) & Engelbracht {\it et al.} (1998) \\
NGC253  & 4-7   & NIR colours~~(screen) & Engelbracht {\it et al.} (1998) \\	
NGC253	& 35	& NIR \& mm recomb. lines(screen)& Puxley {\it et al.}  (1997) \\
NGC253  & 30    & mid-IR line ratios~~(mixed) & Genzel {\it et al.} (1998) \\
NGC3690A/B & 20 & mid-IR line ratios~~(mixed) & Genzel {\it et al.} (1998) \\
NGC4038/9 & 80 & mid-IR line ratios~~(mixed) &  Genzel {\it et al.} (1998)\\
NGC4194 & 0.57($K$) & Br$\alpha$/Br$\gamma$~~(screen) & Mouri {\it et al.} (1997) \\
NGC7469 & 20 & mid-IR line ratios~~(mixed) & Genzel {\it et al.} (1998) \\
NGC7714 & 0.14($K$) & Br$\alpha$/Br$\gamma$~~(screen) & Mouri {\it et al.} (1997) \\
ARP220 & 45 & mid-IR line ratios~~(screen) & Genzel {\it et al.} (1998) \\ \hline
\label{ext}
\end{tabular}
\end{minipage}
\end{table*}

\noindent
where $L_{\rm FIR}$ = $L$(8-1000$\mu$m)$\footnote{The definition of FIR
luminosity varies in literature. Throughout this paper we adopt
$L_{\rm FIR}$=$L$(8-1000$\mu$m)= 4$\pi$$D_{\rm L}^{2}$ $\times$
1.8$\times$10$^{-14}$(13.48$f_{12}$ + 5.16$f_{25}$ + 2.58$f_{60}$ +
$f_{100}$) [W] (Sanders and Mirabel 1996).}$.  The choice of the IMF
cutoffs is not critical here, as a change in the $SFR$ resulting from a
change in the cutoffs is mostly canceled when converting to the SN rate.
This is because the stars exploding as core-collapse SNe are the same
stars which produce the luminosity from which the SFR is estimated.

Applying this $SFR$ estimate to equation (2) we obtain a relation
between the supernova rate and starburst galaxy's FIR luminosity :

\begin{equation}
r_{\rm SN} = \frac{1.1}{\epsilon} \times 10^{-12} L_{\rm FIR} /
L_{\odot}~~yr^{-1}
\end{equation}

For starburst galaxies with high optical depths, $\epsilon$ is about
1.  An example is Arp~220, whose $L_{\rm FIR}=1.4\times10^{12}L_{\odot}$
would therefore imply a supernova rate of 1.5 per year.  However, for
galaxies like NGC~253, M~82, and NGC~4038 the value of $\epsilon$ is smaller
(e.g. Silva {\it et al.} 1998).  Given their $L_{\rm FIR}$ values of
4.2$\times$10$^{10}$L$_{\odot}$, 1.9$\times$10$^{10}$L$_{\odot}$, and
6.8$\times$10$^{10}$L$_{\odot}$ respectively, this implies SN rates exceeding 
0.046, 0.021, and 0.075 per year, consistent with the typical values given in 
Table $\ref{rates}$.

In summary, equ.~(1) based on direct observations of SNRs is
consistent with equ.~(5) and agrees well with the $r_{\rm
SN}$--$L_{\rm FIR}$ relation obtained similarly by Van Buren $\&$
Greenhouse (1994).  It is also similar to the {\it unobscured} type
II+Ib/c SN rate of $(2.5\pm0.3)\times10^{-12}\times$ $L_{\rm FIR}$ /
$L_{\odot}~~yr^{-1}$ obtained by Cappellaro {\it et al.}  (1999).
However, the typical ages of starbursts, 10--100Myr (e.g. Genzel {\it
et al.} 1998), are comparable with the 5--50~Myr lifetimes of the
core-collapse SN progenitors.  Therefore, the relation between the SN
rate and the galaxy's far--IR luminosity depends on the age of the
burst (Fig. 13 in Genzel {\it et al.} 1998), rather than remaining the
same from one starburst to another.  Starburst models indicate ages of
20--30~Myrs for the bursts in M~82 (Efstathiou {\it et al.} 2000) and
NGC~253 (Engelbracht {\it et al.} 1998).  Thus, for a younger or much
older starburst the SN rate would be somewhat smaller.  Equ.~1 will be
used later in this paper to estimate the SN rate for a sample of
starburst galaxies. \\

\section{Extinction Towards Nuclear Supernovae}
The light from supernovae in starburst galaxies is scattered and
absorbed by dust located both locally within the star formation
regions in which the SNe occur, and at greater distances but still
within their host galaxies.  Near-IR recombination lines and colours 
are widely used for deriving the extinction towards starbursts because 
they suffer a lower extinction than optical lines. For example, Engelbracht 
{\it et al.} (1998) find $A_{V}=4-19$ towards the nucleus of NGC~253.
However, even near-IR lines may not probe deeply enough into these
regions.  Hydrogen ionisation rate studies of Puxley {\it et al.}
(1997) based on NIR and mm-wavelength observations, and mid-IR line
studies by Genzel {\it et al.} (1998) indicate considerably higher
extinctions ($A_{V} = 30-35$) towards the NGC~253 nucleus.  In Table
$\ref{ext}$ extinction estimates to galactic nuclei obtained by a
range of techniques are listed for several galaxies. In estimating the
extinction, two extreme models are often considered {\it viz.} a
foreground dust screen and a mixture of stars/gas and dust. These tend
to respectively under- and over-estimate the extinction.
A$_{V}(screen)$ $\sim$ 10 corresponds to $A_{V}(mixed)$ $\geq100$
(Genzel {\it et al.} 1998).  It is also likely that the extinction is
very patchy, adding further to the uncertainty in the true value.  To
reduce these uncertainties, we have carried out a new study of the
extinction in M~82.  This is now described.

\begin{table*}
 \centering
 \begin{minipage}{140mm}
  \caption{Extinction estimates for the SNRs of M~82. The right ascensions and declinations are offsets from 09h51m00s and +69d54m00s.}
\begin{tabular}{@{}llllll@{}} \hline
RA (1950.0) & Dec (1950.0) & $N$(H~I)$_{\rm SNR}$ & $N$(H$_{2}$)$_{\rm LOS}$ & 2$N$(H$_{2}$)$_{\rm LOS}$/ & $A_{V}$ \\             
seconds & arcsec & $\times$10$^{21}$ cm$^{-2}$ & $\times$10$^{21}$ cm$^{-2}$ & $N$(H~I)$_{\rm SNR}$ \\ \hline
39.10 & 57.3 & 3.7 $\pm$ 0.6 & 110 & 59 $\pm$ 31 & 16 $\pm$ 3 \\
39.40 & 56.1 & $>$16 & 123 & $<$15 & $>$71 \\
39.77 & 56.9 & $>$6.5 & 113 & $<$35 & $>$29 \\
40.62 & 56.0 & 9.3 $\pm$ 2 & 204 & 44 $\pm$ 24 & 41 $\pm$ 10 \\
40.67 & 55.1 & 9.5 $\pm$ 0.2 & 160 & 34 $\pm$ 17 & 42 $\pm$ 5 \\
41.30 & 59.6 & 4.7 $\pm$ 1 & 51 & 22 $\pm$ 12 & 21 $\pm$ 5 \\
41.95 & 57.5 & 4.5 $\pm$ 0.1 & 26 & 12 $\pm$ 6 & 20 $\pm$ 2 \\
42.53 & 61.9 & 7.7 $\pm$ 0.4 & 40 & 10 $\pm$ 5 & 34 $\pm$ 4 \\
42.61 & 59.9 & 7.3 $\pm$ 0.8 & 46 & 13 $\pm$ 6 & 32 $\pm$ 5 \\
42.65 & 61.5 & 7.0 $\pm$ 1 & 41 & 12 $\pm$ 6 & 31 $\pm$ 6 \\
43.18 & 58.3 & 7.8 $\pm$ 0.3 & 41 & 11 $\pm$ 5 & 35 $\pm$ 4 \\
43.22 & 61.2 & 9.5 $\pm$ 1 & 42 & 8.8 $\pm$ 5 & 42 $\pm$ 7 \\
43.31 & 59.2 & 5.1 $\pm$ 0.1 & 41 & 16 $\pm$ 8 & 23 $\pm$ 3 \\
43.74 & 62.5 & 5.3 $\pm$ 1 & 40 & 15 $\pm$ 8 & 24 $\pm$ 5 \\
43.81 & 62.8 & $>$13 & 39 & $<$6 & $>$58 \\
44.01 & 59.6 & 4.8 $\pm$ 0.5 & 34 & 14 $\pm$ 7 & 21 $\pm$ 3 \\
44.28 & 59.3 & 3.3 $\pm$ 0.5 & 35 & 21 $\pm$ 11 & 15 $\pm$ 3 \\
44.51 & 58.2 & 6.0 $\pm$ 0.6 & 33 & 11 $\pm$ 6 & 27 $\pm$ 4 \\
44.91 & 61.1 & $>$9.0 & 55 & $<$12 & $>$40 \\
45.17 & 61.2 & 19 $\pm$ 0.4 & 61 & 6.4 $\pm$ 3 & 85 $\pm$ 10 \\
45.24 & 65.2 & $>$22 & 110 & $<$10 & $>$98 \\
45.48 & 64.8 & $>$30 & 136 & $<$9.1 & $>$134 \\
45.74 & 65.2 & 4.2 $\pm$ 2 & 105 & 50 $\pm$ 35 & 19 $\pm$ 9 \\
45.89 & 63.8 & $>$10 & 93 & $<$19 & $>$45 \\
46.52 & 63.9 & 3.9 $\pm$ 2 & 54 & 28 $\pm$ 20 & 17 $\pm$ 9 \\
46.70 & 67.1 & 6.0 $\pm$ 2 & 50 & 17 $\pm$ 10 & 27 $\pm$ 9 \\ \hline
39.64 & 53.4 & $<$6.8 & - & - & $<$30 \\
40.32 & 55.1 & $<$9.5 & - & - & $<$42 \\
42.66 & 55.5 & $<$2.5 & - & - & $<$11 \\
42.80 & 61.2 & $<$2.7 & - & - & $<$12 \\
44.35 & 57.8 & $<$3.3 & - & - & $<$15 \\
44.43 & 61.8 & $<$2.2 & - & - & $<$10 \\
45.42 & 67.4 & $<$1.6 & - & - & $<$7 \\ \hline 
\label{m82snr}
\end{tabular} 
\end{minipage}
\end{table*}

\subsection{Extinction towards the SNRs of M~82}
An estimate of the extinction towards the SNRs in M~82 can be made if
we know the total hydrogen (atomic + molecular) column densities to
these objects.  Wills {\it et al.} (1998) measured 21~cm atomic
hydrogen absorption towards 33 individual SNRs in M~82 from which they
derived H~I column densities.  Wei\ss~{\it et al.} (2000) have used
Large Velocity Gradient (LVG) calculations to obtain molecular
hydrogen (H$_{2}$) column densities from high resolution $^{13}$CO and
C$^{18}$O observations for several lines-of-sight through M~82.  By
interpolation between these values they have provided estimates of the
H$_{2}$ column densities through M~82 at the SNR positions (Wei\ss~private
communication).  In Table $\ref{m82snr}$ (upper section), col.~3 and 4 
respectively, we give the column densities of atomic hydrogen towards the SNRs,
$N$(H~I)$_{\rm SNR}$, and of molecular hydrogen through the whole galaxy
along the line of sight (LOS) to the SNRs, $N$(H$_{2}$)$_{\rm
LOS}$.  Using these values
and assuming a conservative error of $\pm$50$\%$ for the interpolated
N(H$_{2}$) values, we obtain the ratio 2$N$(H$_{2}$)$_{\rm
LOS}$/$N$(H~I)$_{\rm SNR}$ for each SNR.  These are presented in Table
$\ref{m82snr}$, col.~5.  We expect this ratio to have its smallest
values for the SNRs which are located behind most of the molecular gas
along the line of sight {\it i.e.}  the minimum values give us an
indication of the mass ratio of H$_{2}$ to H~I along a given
line-of-sight.

\begin{table*}
 \centering
 \begin{minipage}{180mm}
  \caption{The supernova sample}
  \begin{tabular}{@{}llllllllllllll@{}} \hline
SN & Type & t$_{BVmax}
$\footnote{Epoch in (Julian Days--2400000). Epoch references - 1979C:
Panagia {\it et al.} (1980), 1980K: Dwek {\it et al.} (1983), 1982E:
Maza (1982), 1982R: Muller \& Pizarro (1982), 1983I: Elias {\it et
al.} (1985), 1983N: Panagia {\it et al.} (2000), 1984L: Elias {\it et
al.} (1985), 1985L: Kriss (1985), 1985P: Evans \& Hazelbrook (1985),
1990E: Schmidt {\it et al.} (1993), 1993J: Lewis {\it et al.} (1994),
1994I: Iwamoto {\it et al.} (1994), 1998S: A. Fassia private
communication.}
& Galaxy & distance & $A_{V}$$\footnote{Extinction references - 1979C: Fesen {\it et al.}  (1999),
1980K: Buta {\it et al.} (1982), 1983N: Panagia {\it et al.}  (2000),
1990E: Schmidt {\it et al} (1993), 1993J: Richmond {\it et al.}
(1994), 1994I: Iwamoto {\it et al.} (1994), 1998S: Fassia {\it et al.}
(2000).}$
& $M_{J}$ & $M_{H}$ & $M_{K}$ & $M_{L}$ & $M_{J}$ & $M_{H}$ & $M_{K}$ & $M_{L}$ \\
   &      &           &        & (Mpc)  & & obs.   & obs.  & obs. & obs. & max & max & max & max  \\ \hline
1979C & IIL & 43979 & M 100    & 16.1(1.3) & 0.71(0.1) & --19.67 & --19.84 & --19.95 & --20.38 & --19.9 & --20.1 & --20.2 & --20.4 \\ 
1980K & IIL & 44543 & NGC 6946 & 5.7(0.5)  & 1.1(0.2)  & --18.69 & --18.74 & --18.83 & --19.18 & --18.7 & --18.7 & --18.8 & --18.9 \\
1983N & Ib & 45533 & NGC 5236  & 5.0(0.75)  & 0.85(0.5) & --18.31 & --18.32 & --18.40 & --18.50 & --18.3 & --18.4 & --18.4 & --18.4 \\
1990E & IIP & 47960 & NGC 1035 & 17.1(1.9) & 1.5(0.3)  &  --18.02 &  --17.99 &  --      & -- & --18.0 & --18.0 & -- & -- \\
1993J & IIb & 49095 & M81      & 3.63(0.34) & 0.58(0.4) & --17.95 &  --17.93 &  --18.06 & -- & --17.8 & --17.9 & --17.9 & -- \\
1994I & Ic  & 49451 & M51      & 8.3(1.2)  & 1.4(0.3)  &  --      &  --      &  --17.19 & -- & -- & -- & --17.3 & -- \\
1998S & IIn & 50890 & NGC 3877 & 18.2(2.7) & 0.68(0.1) &  --19.47 &  --19.49 &  --19.48 & -- & --19.7 & --19.6 & --19.6 & --20.4 \\ \hline
$\footnote{The lower section lists SNe for which IR data are not available 
until at least 10~days after maximum $BV$ light.  For these events the peak 
magnitudes were estimated using the template light curve, and are quoted to a 
precision of only one figure after the decimal point. In addition, no 
extinction measurements are available for these SNe and so the 
extinctions were estimated assuming a host galaxy extinction of $A_{V}$
$\sim$ 0.7 plus the Galactic extinction according to Schlegel et
al. (1998).}$
1982E & Ib & 45057 & NGC 1332 & 19.0(2.9) & (0.8)     &  -- & -- & -- & --              & --16.7 & --18.7 & --19.9 & --  \\
1982R & Ib & 45248 & NGC 1187  & 17.5(2.6) & (0.8)     & -- & -- & -- & --              & --17.6 & --17.5 & --17.7 & --19.3  \\
1983I & Ic & 45443 & NGC 4051  & 18.2(2.7) & (0.7)     & -- & -- & -- & --              & --18.2 & --18.7 & --18.4 & --  \\ 
1984L & Ib & 45943 & NGC 991   & 20.1(3.0) & (0.8)     & -- & -- & -- & --              & --19.0 & --19.4 & --19.1 & --   \\
1985L & IIL & 46227 & NGC 5033 & 20.0(3.0) & (0.7)     & -- & -- & -- & --              & --20.3 & --20.2 & --20.1 & --19.7 \\
1985P & IIP & 46349 & NGC 1433 & 12.4(1.9) & (0.7)     & -- & -- & -- & --              & --19.4 & --19.4 & --19.3 & --  \\ \hline
\label{magcc}
\end{tabular}
\end{minipage}
\end{table*}

Uncertainties in the ratios are due to errors in the $N$(H~I) values,
the CO data entering the LVG calculations and the interpolations to
the SNR positions.  In order to make better use of the column density
data, we follow the philosophy of the V/V$_{max}$ test (Schmidt
1968).  In this we assume that 2$N$(H$_{2}$)$_{\rm
LOS}$/$N$(H~I)$_{\rm LOS}$ has a single value for any line of sight
through M~82.  In addition, we assume that the SNRs are randomly
distributed throughout M~82, and so we expect to have :

\begin{equation} 
< \frac{N(H~I)_{\rm SNR}}{N(H~I)_{\rm LOS}} > = \frac{1}{2}   
\end{equation}

Applying the test using the 19 sets of H~I (excluding the upper and
lower limits) and H$_2$ column densities (table $\ref{m82snr}$), we
find an H$_{2}$ to H~I mass ratio of 7.4 $\pm$ 1.0.  However, it should be 
noted that the H$_{2}$ column densities are model dependent introducing 
further uncertainty in the mass ratio.  Nevertheless, our result is 
consistent with the lowest values listed in table~3, but has a higher 
precision.  It is also in good agreement with the mass ratios of 
7--10 suggested by Crutcher {\it et al.} (1978) and Yun {\it et al.} 
(1993) for the centre of M~82.

\epsfxsize=17cm
\begin{figure*}
 \vspace{0.5cm}
 \epsfbox{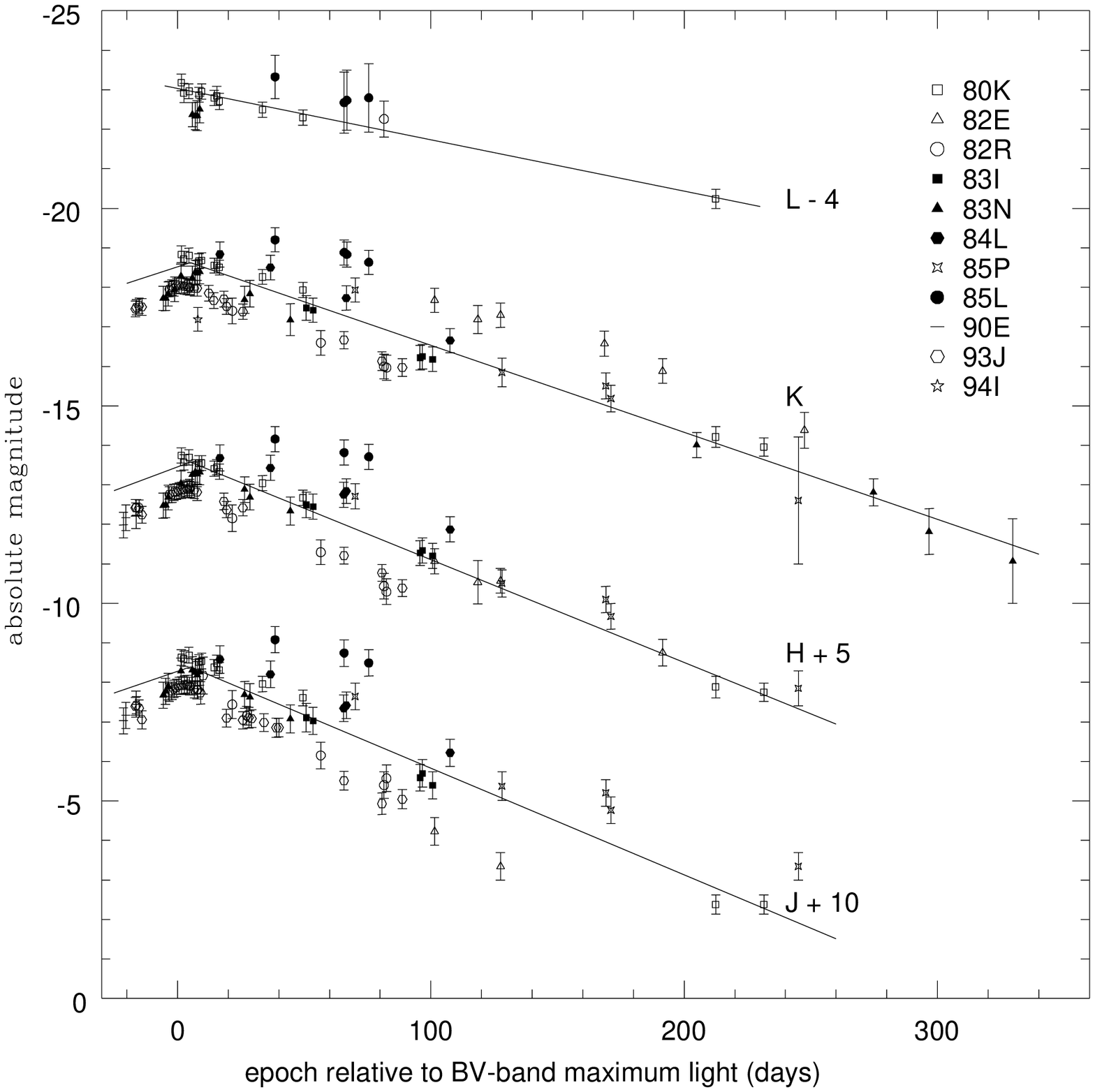}
 \caption{$JHKL$-band light curves and templates for ``ordinary''
core-collapse SNe.  For clarity, the $J$, $H$, and $L$-band data and
templates have been vertically displaced by +10, +5, and --4 magnitudes
respectively.  The horizontal positions of the individual light curves
were set such that 0~days corresponds to $BV$-maximum (see text) in
each case.  The error bars give the combined uncertainties in photometry,
extinction and distance.}
 \label{combinedII}
\end{figure*}

\epsfxsize=17cm
\begin{figure*}
 \vspace{0.5cm} 
\epsfbox{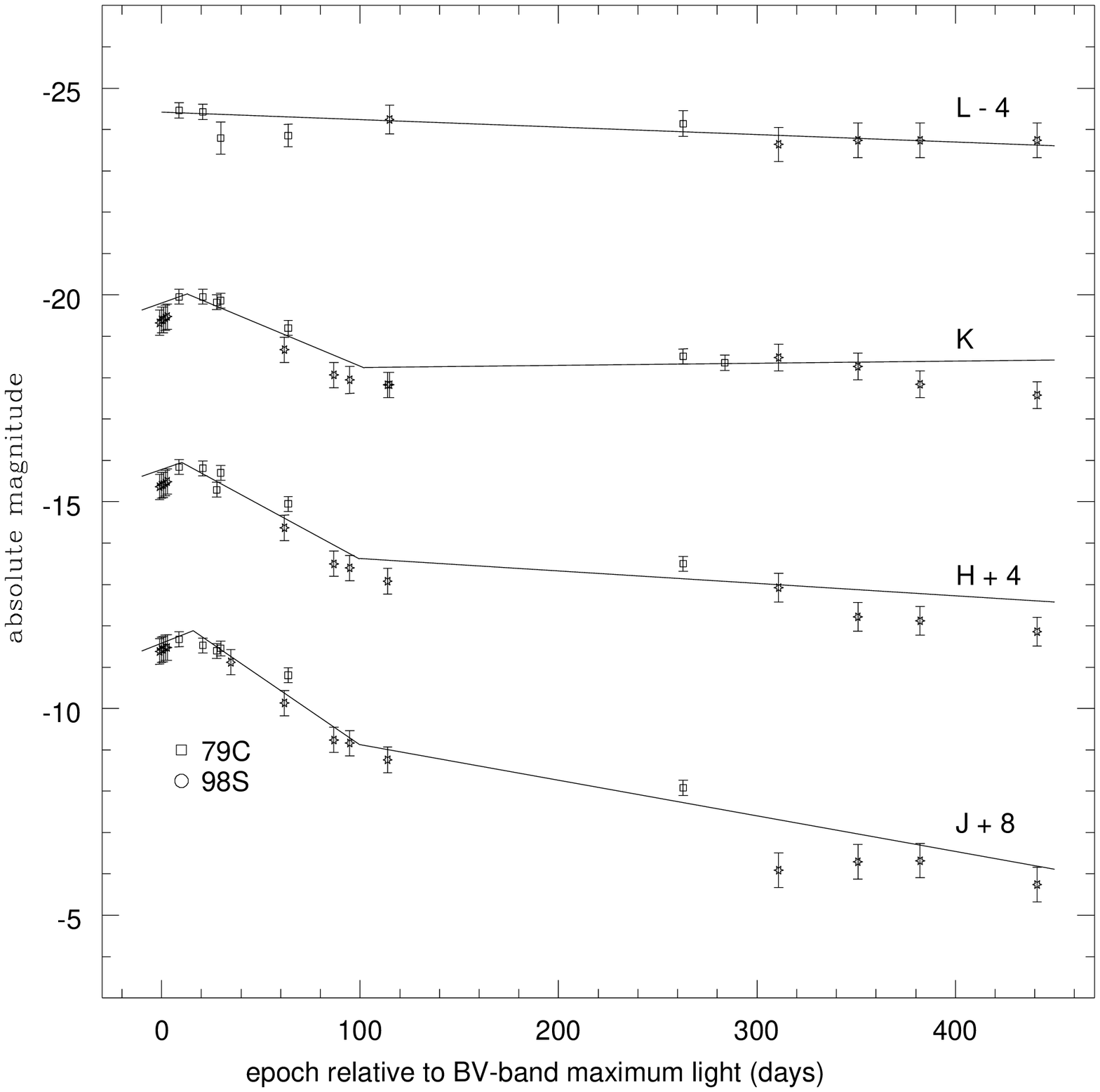} 
\caption{$JHKL$-band light curves and templates for
``slowly-declining'' core-collapse SNe.  For clarity, the $J$, $H$,
and $L$-band data and templates have been vertically displaced by +8,
+4, and --4 magnitudes respectively.  The horizontal positions of the
individual light curves were set such that 0~days corresponds to
$BV$-maximum (see text) in each case.  The error bars give the combined 
uncertainties in photometry, extinction and distance.}
\label{combinedIIn}
\end{figure*}

The extinction towards the M~82 SNRs was estimated adopting $A_{V}$ =
0.53 $\times$ 10$^{-21}$ ($N$(H~I) + 2$N$(H$_{2}$)) cm$^{-2}$ (Bohlin
{\it et al.} 1978).  This assumes that the dust properties and
dust/gas ratio are the same for M~82 as for the Milky Way Galaxy.  For
each SNR, an estimate of A$_{V}$ was obtained by adopting 7.4 for the
2$N$(H$_{2}$)/$N$(H~I) ratio and using the $N$(H~I) values.
The extinctions are listed in Table~3, col.~6.  The uncertainties arise from
errors in the $N$(H~I) measurements and in the adopted 2$N$(H$_{2}$)/$N$(H~I) 
ratio.  We note that the SNRs with the highest extinctions are located 
within the two CO lobes on the edges of the galaxy.  It has been suggested 
that these lobes are associated with the ends of an edge-on ring structure 
or bar structure (Neiniger {\it et al.} 1998).  Considering only those 19 
SNRs for which we have $N$(H~I) values (as against lower or upper limits), we
obtain a weighted mean extinction of A$_{V}$ = 24 ($\sigma$$\sim$9).  However, 
M~82 is an almost edge-on galaxy and so, in general, the column densities 
will be considerably greater than for a face-on starburst galaxy.  
Incorporation of our M~82 extinction analysis into the feasibility study will 
be described in Section 6.1.

\section{The IR light curves of core-collapse Supernovae}
In order to investigate realistically the feasibility of discovering
nuclear SN, we need to ascertain the IR luminosity and evolution of
core-collapse SNe.  To this end we have assembled near-IR photometric
data for a total of 14 events, some of which has never been published.
This constitutes the most complete database of core-collapse SN IR
light curves to date.  However, one event was SN~1987A, and although
coverage of its IR light curves is unparalleled, they were highly
atypical in luminosity and shape. We have therefore excluded this
event.  The photometric data for the remaining 13 SNe are listed in
Appendix~B.  In Table $\ref{magcc}$ we list other information about
these SNe.  We have used these events to derive absolute magnitude
``template'' light curves to represent the typical core-collapse SN 
evolution in the IR.

Distances to the host galaxies are listed in Col.~5.  For
SN~1979C (Ferrarese {\it et al.}  1996) and SN~1993J (Freedman {\it et
al.}  1994) these were obtained from HST Cepheid observations.  For
SN~1980K, Eastman {\it et al.} (1996), Buta {\it et al.} (1982), and
Tully (1988) give similar distances using, respectively, the expanding
photosphere method, photometric distance indicators, and recession
velocities corrected for the expansion of the Local Super Cluster and
the infall towards the Virgo Cluster.  We adopted the average of these
three values.  For SN~1990E, we adopted the average of the similar
distances given by Eastman {\it et al.} (1996), and Tully (1988).  For
the remaining nine SNe, we used the distances given by Tully
(1988)\footnote{All the distances given by Tully (1988) have been
scaled to H$_{\rm 0}$ = 70 kms$^{-1}$ Mpc$^{-1}$.}.  Estimates of the 
errors in the distances are given in parentheses.  

For seven events (SNe 1979C, 80K, 83N, 90E, 93J, 94I, 98S) photometry
is available for at least one of the $JHKL$ bands less than
$\sim$10~days after maximum light in the $BV$-band $\footnote{The
actual peak magnitude of core-collapse SNe is normally missed. ``$BV$
maximum light'' in this context usually means the brightest observed
magnitude on the already-declining light curve.  However, in the case
of SN~1983N, the $B$-band light curve was followed through the true
maximum.  In this paper we adopt the time of maximum light in the
$BV$-band as epoch 0~days.}$.  Extinction estimates for these SNe are
also available from the literature, and these are listed in Col.~6.
These values were converted to $A_J$, $A_H$, $A_K$, and $A_L$ assuming
the galactic reddening law given by Rieke $\&$ Lebofsky (1985).
Absolute intrinsic magnitudes near epoch 0~days were then determined
by subtracting the distance moduli from the apparent magnitudes and
correcting for extinction.  These magnitudes are listed in Cols.~7-10.
The other six events (SNe~1980E, 80R, 83I, 84L, 85L, 85P) were not observed 
until an epoch of later than 10~days.  Moreover, no extinction estimates were 
available for these events.
Therefore, the extinctions were estimated by adopting a typical host
galaxy extinction of $A_V=0.7$ and adding to this the appropriate
Galactic extinction value given in Schlegel et al. (1998).  The
uncertainty in $A_V$ is therefore quite large - we estimate $\pm$
0.5. The peak magnitudes for these SNe given in Table $\ref{magcc}$
(cols.~11-14) are only estimates, obtained by extrapolating back to
epoch +5~days for ``ordinary'' events and to $\sim$+10~days for
``slowly-declining'' events using the light curve templates described
below.

Among the 13 events used to determine the absolute magnitude template
light curves, we identify two classes of behaviour.  In one class we
have 4 supernovae (including types~IIP, IIL, Ib) which decline
linearly for at least $\sim$200~days after epoch 0~days.  We shall
refer to these as ``ordinary'' events. This class includes SNe~1980K,
82E, 83N and 85P.  In the other class we have 2 ``slowly-declining''
supernovae, comprising the type~IIL SN~1979C and the type~IIn SN~1998S.
From about 0~days to 100~days the $JHK$ light curves of this class
decline at a rate similar to that seen in the ordinary light curves.
However, beyond $\sim$100~days a much slower decline in $J$ and $H$
occurs, together with virtually no decline in $K$, persisting for over
450~days.  Moreover, in the $L$ band there is negligible decline from
soon after the explosion to beyond 450~days.  Recent observations show
that the slowly-declining behaviour has persisted in SN~1998S for over
700~days. The remaining 7 events (SNe~1982R, 83I, 84L, 85L, 90E, 93J,
94I) all show a similar decline rate during the first 100~days to that
seen in the ordinary and slowly-declining classes.  They include types
IIP, IIL, IIb and Ib/c.  However, as they were not observed in the IR
much beyond 100~days we are unable to distinguish which of the two
classes, if either, they should be assigned to.  We therefore,
conservatively, assign them to the ordinary class.

We constructed absolute magnitude template light curves for the two
classes. This was done as follows.  Inspection of the ordinary
individual $JHK$ light curves suggested that two-component linear
templates might provide a reasonable description of the data.  We
therefore fitted functions of this form to the entire ordinary
absolute magnitude dataset in each band.  Epoch +5~days was selected
by visual inspection as a plausible intersection point for the two
components.  The template fit for the data after +5~days was then
obtained in two stages.  In the first step, the best-fitting slope was
determined by iteratively adjusting the vertical position of the
individual SN light curves.  The reason we followed this procedure was
to minimise bias in the slope which would otherwise be introduced by
differences in temporal coverage for different SN events having
different peak magnitudes.  As we were only interested in the slopes
at this stage, only the photometric errors were included in the fits.
Reduced $\chi^2$ values of 20, 10, 6 were obtained in $JHK$
respectively.  The larger-than-unity values of the reduced $\chi^2$s 
obtained indicates some intrinsic variations in light curve shapes between 
SN events.  In the second step, the absolute peak magnitudes for
the templates were obtained. For each band, the post-day~+5 template
slope was fixed at the value obtained in the procedure just described.
The templates were then compared with the original, unshifted data points
of individual supernovae.  The vertical position of the post-day~+5 
lightcurve of each individual supernova was adjusted to obtain the best 
fit to the points {\it i.e.} the only free parameter was the absolute 
magnitude position of the lightcurve.  The absolute peak magnitude of
the template was then determined by taking a weighted average of the 
individual supernova peak magnitudes (Table~4, cols.~11-14).  The weights 
for the events were obtained by combining the error in the apparent peak 
magnitude obtained in the fit with the errors in the distance modulus 
($\pm$0.2--0.3~mag) and in the extinction corresponding to uncertainties 
in $A_V$ of $\pm$0.1--0.5 (see Table~$\ref{magcc}$).  Turning now to the 
pre-+5~day template component, the +5~day magnitude of this component was set 
to match that of the post-+5~day component.  The slope of the pre-+5~day 
component was then varied to provide the best fit.  The fits were carried out 
after shifting the data points vertically so that the peak magnitudes of the 
individual events and the template were the same.  Only the photometric 
errors were included in the fit.  Reduced $\chi^2$s of 50, 86, and 63 in 
$JHK$ respectively were obtained, indicating a considerable 
variation in the pre-maximum lightcurve slopes.  For the $L$ data, a single 
linear fit provided an adequate description.  As with the $JHK$ bands, 
the best slope was found by iteratively adjusting the vertical
position of the individual SN light curves.  This yielded a reduced 
$\chi^2$ of 1.0.  Again, the vertical position of the template was determined
by a weighted average of the absolute magnitudes of the individual events
obtained in the same way as before.

\begin{table}
 \begin{minipage}{90mm}
  \caption{Template light curve of ordinary SNe}
  \begin{tabular}{@{}llllllllllllll@{}} \hline
           & J     & H     & K     & L      \\ \hline 
$\delta_{1}$  & --0.022 & --0.024 & --0.021  & -- \\
M(5)  & --18.39 & --18.58 & --18.62 & --18.97 \\
$\delta_{2}$  & 0.027 & 0.026 & 0.022 & 0.013 \\ \hline
$\sigma_{y}$ & 0.86 & 0.71 & 0.82 & 0.41 \\ \hline
\label{ordinary}
\end{tabular}
\end{minipage}
\end{table}

\begin{table}
 \begin{minipage}{90mm}
  \caption{Template light curve of slowly declining SNe}
  \begin{tabular}{@{}llllllllllllll@{}} \hline
           & J     & H     & K     & L      \\ \hline 
$\delta_{1}$  & -0.019 & -0.017 & -0.017 & 0.0018   \\
M(1)       & --19.88  & --19.95  & --20.02  & --20.42 \\
t$_{1}$	   & 16 & 10 & 13 & 0 \\	
$\delta_{2}$  & 0.033 & 0.026 & 0.020 & 0.0018  \\
M(2)       & --17.12 & --17.63 & --18.24 & - \\     
t$_{2}$    & 100 & 100   & 102  & - \\
$\delta_{3}$  & 0.0086 & 0.0030 & -0.00053 & 0.0018 \\ \hline
$\sigma_{y}$ & 0.42 & 0.27 & 0.46 & 0.96 \\ \hline
\label{slow}
\end{tabular}
\end{minipage}
\end{table}

The $JHKL$ ordinary template parameters are given in Table
$\ref{ordinary}$.  In a given band, M(5) is the absolute magnitude of
the template peak at +5~days.  $\delta_{1}$ and $\delta_{2}$ are the light
curve slopes before and after epoch +5~days.  The decline rates ranged
from 0.027 mags/day in $J$ to 0.013 mags/day in $L$.  In the bottom
row we give the weighted dispersion, $\sigma_{y}$, of the individual supernova
peak magnitudes about the peak of the template.  The ordinary templates 
and individual data points are shown in Figure~$\ref{combinedII}$, where 
the horizontal position (epoch) is with respect to the epoch of 
$BV$-maximum = 0~days (see footnote).  The error bars on the data points 
are the combined errors in photometry, extinction, and distance.

For the two slowly-declining SNe, we adopted three-component linear
templates in $JHK$, with a single-component linear template in $L$.
Guided by visual inspection, we divided the data into three eras
corresponding to pre-+10~days, +10--+100~days, and post~+100~days.
Linear fits were then carried out for each component.  Only the errors
in photometry were included.  The best fitting slope for the epoch
between 10 and 100 days was obtained again by iteratively adjusting
the vertical positions of the individual events.  This slope was
then used for estimating the absolute peak magnitudes of both  
events, and thus the vertical position of the template as before.
The data points of the individual events were then shifted vertically 
so that their peak magnitudes matched the one of the template.
This allowed us to perform linear fits for the pre-+10~day, and
post-+100~day components.  The magnitudes of the points of 
intersection ({\it i.e.} the template peak and inflection point) 
were derived from the fits.  Reduced $\chi^2$s were 41, 29, and 2 
for +10--+100~days and 15, 
11, and 14 for post-+100~days in $JHK$ respectively.  The value for the
single-component $L$-band fit was 2.  For the pre-+10~days, the
reduced $\chi^2$s were 4, 10, and 5 in $JHK$ respectively.
The template parameters are shown in Table $\ref{slow}$.  M(1) and t$_{1}$
give, respectively, the absolute magnitude and epoch of the template
peak, while M(2) and t$_{2}$ give the absolute magnitude and epoch of
the inflection.  $\delta_{1}$, $\delta_{2}$ and $\delta_{3}$ give the
respective slopes for the three eras.  The bottom row gives the
dispersion, $\sigma_{y}$, of the post-+10~day points about the
template.  The templates and individual data points (with the photometry
errors) are shown in Figure~$\ref{combinedIIn}$ where, as before, the 
horizontal position (epoch) is with respect to the epoch of $BV$-maximum = 
0~days (see footnote).  As already indicated, we find that up to about 
day~+100, the decline rates are the same for both classes, to within the 
errors.  However, the defining characteristic of the slow-decliners is the
slope after +100~days.  For this era, we obtain slopes of 0.0086,
0.0030, $\sim$0, and 0.0018~mags/day in the $J$, $H$, $K$, and
$L$-band respectively.  However, our study has revealed an important
additional characteristic of the slow-decliners {\it viz.} that they
are significantly more luminous than the ordinary class, even at early
times.  Around maximum light SNe in this class are $\sim$1.5~mags
brighter in $JHKL$.  However, it should be remembered that we have 
identified only 2 slow-decliners in our sample.  Presumably, the 
slowly-declining late-time light curve of this class, and perhaps also 
the greater early-time luminosity, is being powered by conversion of the 
SN kinetic energy into IR radiation through shock interaction with the 
circumstellar medium and heating of dust local to the supernova
(see {\it e.g.} Fassia {\it et al.} 2000a, Fassia {\it et al.} 2000b).
  
Because of the reduced extinction, and the slower light curve decline 
rates compared with the $J$- and $H$-bands, the $K$- and $L$-bands are
superior for the purpose of a nuclear SN search.  Below we examine the 
feasibility of detecting such SNe using repeated $K$-band imaging of a 
sample of nearby starburst galaxies. 

\section{A technique for supernova detection}
Having estimated the intrinsic rate of core-collapse supernova
explosions (Sect. 2), determined the typical range of extinctions
towards SNe in starburst galaxies (Sect. 3), and established the
magnitude and shape of $K$-band SN light curves (Sect. 4) the next
stage is to examine the efficiency with which we might detect 
SNe in galaxy nuclear regions.

\subsection{$K$-band imaging of NGC~5962}
We used an image pair of the IR-luminous galaxy NGC~5962 to study the
efficiency of SN detection in the galaxy nuclear regions.  NGC~5952 is
a star-forming galaxy at a distance of 34 Mpc (Tully 1988) with a far-IR 
luminosity similar to the prototypical starburst galaxy M~82 ($L_{\rm FIR}$ = 
3.8$\times$10$^{10}$ L$_{\odot}$).

\begin{figure*}
 \vspace{-2.5cm}
 \hspace{-0cm}
 \vbox to90mm{\vfil
 \epsfbox{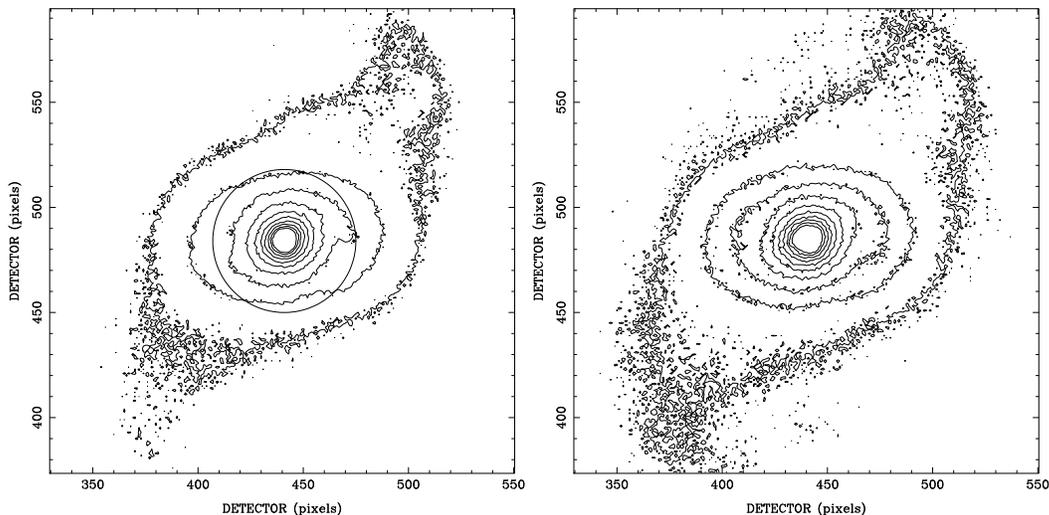}
 \vspace{-15.5cm}
 \hspace{+5cm}
\caption{$K$-band images of NGC 5962 obtained with the UFTI camera at UKIRT
(North is up and East to the left). The plate scale is 0.091'' per
pixel.  The April~4 image (seeing FWHM=0.6'') is on the left and the
April~5 image (seeing FWHM=1.0'') is on the right. The circle in the
left hand image is of radius 500~pc ($\sim$3''). The contour interval
is 40 counts.  The lowest and highest contours are 50 and 410 counts
which correspond to +15.9 and +13.6 mags per square arcsec,
respectively, in the left hand image, and +15.7 and +13.5 mags per
square arcsec, respectively, in the right hand image.}
\label{fig3} 
\vfil}
\end{figure*}

State-of-the-art $K$-band images of NGC~5962 were obtained by
A.~Fassia, M.~Hernandez and T.~Geballe using UFTI (0.091''/pixel) at 
UKIRT$\footnote{The 
United Kingdom Infrared Telescope is operated by the Joint Astronomy Centre 
on behalf of the U.K. Particle Physics and Astronomy Research Council.}$.  
Repeat images 
were obtained on 1999 April 4 and 5. The seeing on
the first night was about 0.6'' and about 1.0'' on the second night.
Each image comprised a mosaic pattern of 5 individual frames with 10''
or 11'' offsets.  The total integration time for a complete image was
400--600~sec. For each of the observations a sky flat was observed
with an equal exposure time and dither pattern.  In addition a 5
$\times$ 60 sec exposure sky flat was obtained at the beginning of
both nights.  During the data reduction it was found that the best
result for the April 4 image was achieved by using the contemporary
sky flat for flat fielding.  For the April 5 image, a better result was
achieved with the sky flat taken at the start of the night.  The
images were calibrated using the UKIRT faint standards FS21 and FS25.
The images are shown in Figure~$\ref{fig3}$.

\subsection{Supernova detection}
Supernovae occurring in the nuclear regions of galaxies can be sought
by comparing two images taken at different times. A rough comparison
can be made by examining the images by eye and varying the contrast
levels.  A more efficient approach is by ``blinking'' the two frames.
It is likely that the most sensitive search procedure is to carry out
computerised subtraction of one image from the other to reveal very small
differences.  This is the method used in the feasibility study
described here.  However, for this to be practical it is necessary to
first correct for the effects of differing atmospheric conditions
(seeing and transparency), exposure times, focus positions and
telescope guiding accuracy between the two images.  Ideally, the
images should all be taken with the same instrument.

\subsubsection{Image Matching}
Before attempting to subtract one image from another, they must first be
aligned accurately.  The shifts in $x$ and $y$ are obtainable from the
positions of the galactic nucleus and any bright point sources within
the images.  Due to the small pixel size of UFTI, the alignment did not 
need to be performed to a precision of better than 1 pixel.

%FIGURE 3
\epsfxsize=17cm

%FIGURE 4
\epsfxsize=17cm
\begin{figure*}
 \vspace{-0.5cm}
 \hspace{-0cm}
 \vbox to90mm{\vfil
 \epsfbox{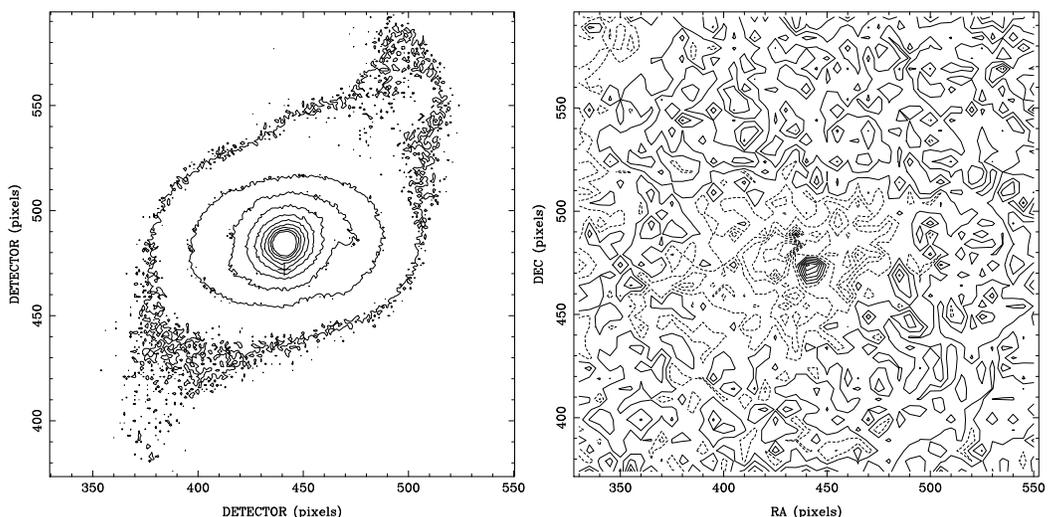}
 \vspace{-15.5cm}
 \hspace{+5cm}
\caption{$K$-band image of NGC 5962 (seeing FWHM=0.6'') with a simulated
SN (cross) of $m_{K}$=17 at (441,472) (left) and after image
subtraction (right). The plate scale is 0.091'' per pixel.  The
contour interval in the subtracted image is 1 count, and the zero
contour is at 2 counts, corresponding to +19.3 mags per square arcsec.
The residual image (right) has been binned by $\times$5 in the $x$ and
$y$ directions.  } 
\label{fig4} 
\vfil}
\vspace{+1cm}
\end{figure*}

A more difficult problem is image mismatch owing to differences in
seeing and photometric conditions between the two observations.
Initially, we attempted to deal with this by matching the point spread
functions (PSFs) and the intensity levels of the two images using the
standard {\tt IRAF}~\footnote{{\tt IRAF} is distributed by the
National Optical Astronomy Observatories, which are operated by the
Association of Universities for Research in Astronomy, Inc. under
contract with the National Science Foundation.} image matching package
{\tt IMMATCH}.  However, this was found to give unsatisfactory results
in regions around bright stars and galactic nuclei.  We therefore
turned to the more sophisticated Optimal Image Subtraction method of
Alard \& Lupton (1998) which they developed for analysis of
microlensing survey images.  It derives an optimal kernel solution
from least squares analysis of the data.  The method has the advantage
that it does not require any bright isolated stars to determine the
kernel but can be used on any portion of an image with high enough
signal-to-noise ratio.  In our feasibility study we used a new version
of the method (Alard 2000) which was developed to process non-crowded
field images.  In this version, instead of using all the image pixels,
only the regions around selected bright, but not necessarily stellar, 
objects are used to find the kernel since in a non-crowded image most 
of the pixels do not include
any useful information for deriving the kernel solution.  Among the
bright objects we must include the innermost region of the galactic
nucleus if we are to obtain a satisfactory subtraction result for the
nuclear regions.  Unfortunately, this makes the
detection of supernovae falling on those innermost pixels impossible.
In the images of NGC~5962 the size of the nuclear region involved in
the kernel solution was 16 $\times$ 16 pixels making detection of SNe
falling within the innermost $\sim$0.7'' impossible.  However, we
estimate that only $\sim$20\% of the supernovae would occur within this 
region (see Section~6).  Both the nucleus and also regions around four 
brightest point sources ($m$$_{\rm K}$ $>$ 16) outside the galaxy's 
circumnuclear regions were used for the fit.  In order to get the 
best results, all the 
regions used for the fit must be within the common portion of the mosaic 
image.  Here, all the regions used for the fitting procedure were within 
30'' of the nucleus and are thus easily fitted within a single field of 
view of a typical infrared imager.

\subsubsection{Testing the Optimal Image Subtraction Method for SN detection}
We used the {\tt IRAF} package {\tt MKOBJECTS} to add artificial
SNe to the galaxy images.  To set the intensity profile of the
artificial SNe, an elliptical Moffat profile with a similar position
angle, ellipticity and beta parameter to the real stars in the image
was used.  This was set to correspond to the
better-seeing image ($\sim$0.6'') and the artificial supernovae were
added to this image.  However, since the better-seeing image is
convolved to match the poorer-seeing image before the image
subtraction, the size of the supernova seeing disk in the residual
image is $\sim$1''.  An example of supernova detection is presented 
in Figure $\ref{fig4}$.  In Figure $\ref{fig4}$(left) an artificial supernova
of $m_{K}=+17.0$ has been placed 1.1'' (180~pc) south of the centre of
NGC~5962 in the better-seeing image, indicated by the cross. In the
residual image (Figure $\ref{fig4}$(right)) the supernova is clearly
visible.  

\begin{figure*}
 \vspace{-0.5cm}
 \hspace{+10cm}
\vbox to90mm{\vfil
 \epsfbox{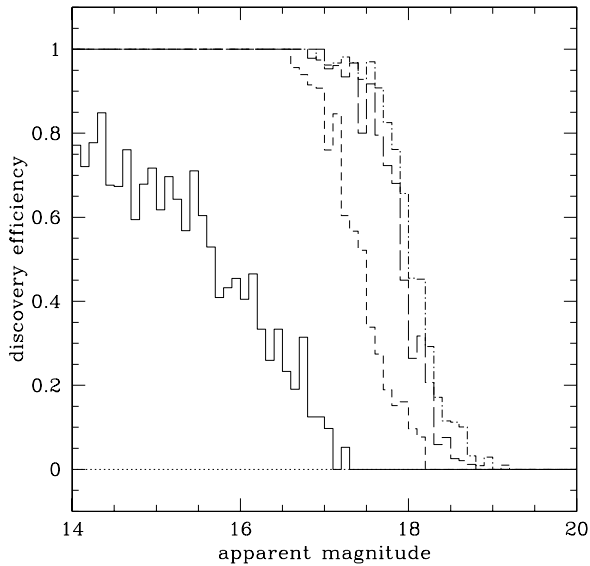}
 \vspace{-0.5cm}
 \hspace{+0cm}
 \caption{The fraction of simulated SNe recovered within different annulae in
 the $K$-image of NGC~5962 as a function of the SN apparent magnitude.  The 
 dotted, solid, short dashed, long dashed, and dashed-dotted lines correspond 
 to the 0-100, 100-200, 200-300, 300-400,  and 400-500pc annulae respectively.}
\label{efficiency}
\vfil}
 \vspace{+0cm}
\end{figure*}

The occurrence of 18000 supernovae was simulated, distributed randomly
within the innermost 500pc radius of NGC~5962.  For each simulation,
subtraction was performed using the procedures described above.  The
residual images were then sought for point sources using the {\tt
IRAF} {\tt DAOPHOT} package.  The detection threshold was set at 5
times the background noise.  The supernova detection efficiency at
different distances from the centre of NGC~5962 is presented in Figure
$\ref{efficiency}$ for a range of apparent magnitudes.  It shows the
probability of detecting SNe of different apparent magnitudes and
positions.  Only in the 0-100~pc region could no supernovae be detected
by this method.

\epsfxsize=8cm

\section{Feasibility of searching for supernovae in the nuclear starburst 
regions of galaxies}
The SN detection efficiency derived above is used here to determine
the number of nuclear supernovae which would be discovered in a
practical survey.

\subsection{Monte Carlo Simulations of the appearance of Supernovae}
We examined the detectability of supernovae in the nuclear regions of
starburst galaxies as follows.  We simulated the range of absolute
peak magnitudes by applying Gaussian distributions to the two types of
template light curve.  Average peak magnitudes of $K$ = --18.6 and
$K$ = --20.0, and dispersions of $\sigma_{K}$ = 0.8 and $\sigma_{K}$
= 0.5 were adopted for the ordinary and slowly-declining SNe
respectively.  Since 1987, the fraction of core-collapse supernovae
exploding as type~IIn events listed in the Asiago Supernova Catalogue
(Barbon {\it et al.}  1999) is 14$\%$.  However, the intrinsic rate of
type~IIn SNe may be smaller since their higher intrinsic brightness
compared to the ``ordinary'' SNe means that they are more likely to be
discovered.  According to Cappellaro {\it et al.} (1997) they
constitute only 2--5$\%$ of all type~II SNe.  However, as the
proportion of slow-decliners in the nuclear starburst regions is
uncertain we carried out the simulation for slow-decliner fractions
ranging between 0 and 60$\%$.  The epoch of observation was allowed to
vary randomly between --20 and +340~days relative to the $K$ maximum
light.  The absolute magnitude for each of the events was then assigned 
according to the template light curves (Section~4).

\epsfxsize=17cm
\begin{figure*}
 \vspace{-1.0cm}
 \hspace{-0cm}
\vbox to90mm{\vfil
 \epsfbox{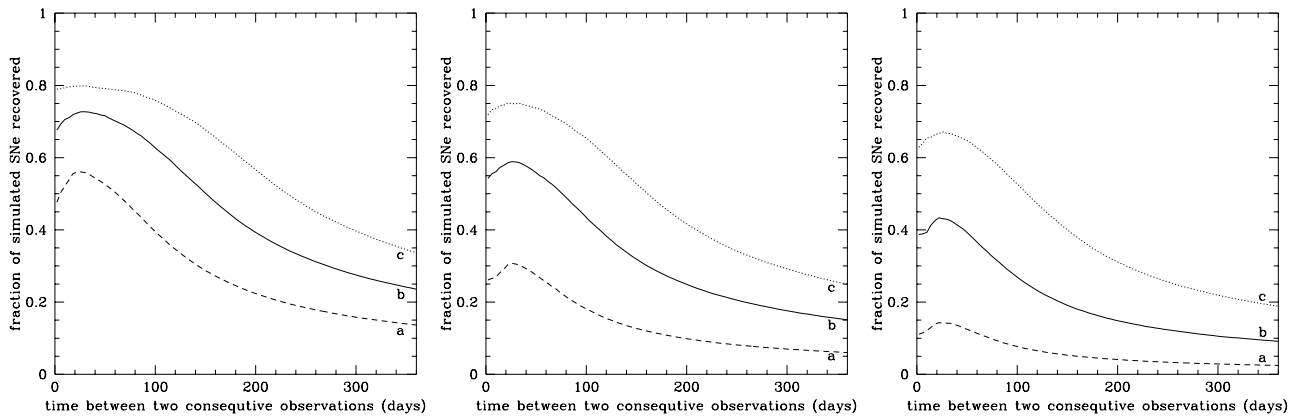}
 \vspace{-18cm}
 \hspace{+10cm}
 \caption{The fraction of simulated SNe recovered within the innermost
 500~pc in the $K$-image of NGC~5962 for different time intervals
 between observations.  Figures (left), (middle) and (right) correspond to
 simulated galaxy distances of $d$ = 20, 30 $\&$ 40~Mpc respectively.
 In each figure, the (a) dashed, (b) solid and (c) dotted lines correspond to
 $A_{V}$ = 18, 27 \& 36 respectively.}
\label{fig7}
\vfil}
 \vspace{-1cm}
\end{figure*}
\epsfxsize=17cm
\begin{figure*}
 \vspace{-1.5cm}
 \hspace{+10cm}
\vbox to90mm{\vfil
 \epsfbox{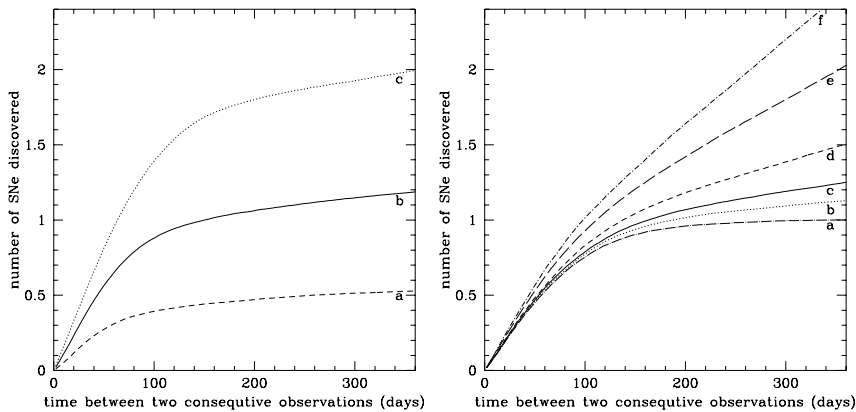}
 \vspace{-18cm}
 \hspace{+0cm}
 \caption{The total number of supernovae discoverable in a single
 observation of the whole galaxy sample (Table~A1) as a function of
 the length of time since the previous observation, $t_{intv}$ .  In
 the LH plot, the (a) dashed, (b) solid and (c) dotted lines correspond to 
 $A_{V}$ = 36, 27 \& 18 respectively, with a 5$\%$ fraction of slow-decliners.
 In the RH plot, we have adopted an M~82-like extinction distribution,
 and the lines a, b, c, d, e, f correspond to different intrinsic fractions of
 slowly-declining events viz. 0, 0.05, 0.10, 0.20, 0.40, and 0.60 respectively
 (see Section 6.2).}
 \label{fig8} \vfil}
 \vspace{+0cm}
\end{figure*}

The absolute magnitudes were then scaled by the appropriate
extinctions and distance moduli.  To simulate the effects of
extinction, we carried out the study for three fixed extinction values
{\it viz.} $A_{K}$ = 2, 3, 4 ($A_{V}$ = 18, 27, 36).  We also carried
out the simulation for a specific distribution of extinction values
towards the nuclear SNe, by adopting the extinction distribution
towards the SNRs in M~82.  This was done by selecting an extinction for
each simulated event randomly from the extinctions listed in Table~3, 
col.~6.  In the case of the extinction lower limits the SN was assumed to
be totally obscured in the $K$-band regardless of its magnitude, whereas the 
extinction upper limits listed in Table~3 (lower section) were considered as 
actual extinction values.

We simulated the occurrence of supernovae in galaxies at distances of
20, 30 and 40~Mpc, and also carried out the simulation for a specific
distribution of distances according to the galaxy sample described in
Appendix~A.  This was done by selecting the distances for the simulated
events from a random distribution in which each of the distances had
a weight according to the corresponding galaxy's far-IR luminosity 
{\it i.e.} its SN rate.  As the distance increases, not only does the supernova
become fainter but also, for conservatively poor seeing of 1'', the
ability to spatially resolve any SN against the nuclear region rapidly
diminishes.  1'' at 50~Mpc corresponds to 240~pc, which is already
half of our search radius.  Therefore, we considered the detection of
SNe within 45~Mpc only.  In considering different host galaxy
distances, we only varied the SN magnitudes.  The actual NGC~5962
images were unchanged.  For galaxies at distances less than 34~Mpc
(the distance to NGC~5962) this is clearly a conservative approach
because of the larger linear size of the galaxy nuclear profile
compared to the size of the seeing disk.  However, for galaxies more
distant than NGC~5962 the reverse is true.  The linear region over
which the SN can be spatially resolved against the nucleus diminishes.
The poorer seeing image of the NGC~5962 pair was 1.0''. Thus, to
achieve comparable resolution for a galaxy at 45~Mpc, we would
need the seeing of the poorer image to be 0.75''.  We therefore
assumed that in the survey, the best seeing nights would be devoted to
the most distant galaxies of the sample.

The intrinsic rate of core-collapse SNe in the nuclear regions of the
sample galaxies was scaled according to the galaxy FIR luminosities,
$L_{\rm FIR}$ (see Appendix~A).  For the relation between the
core-collapse SN rate and $L_{\rm FIR}$ we adopted the average of that
found in NGC~253, M~82, and NGC~4038/39 {\it viz.}  $r_{\rm SN}$ /
$L_{\rm FIR}$ = 2.7 $\times$ 10$^{-12}$ / L$_{\odot}~~yr^{-1}$.  The
supernova location was based on the observation by several authors
that the $K$-band continuum in a starburst system ({\it e.g.} NGC~253,
Engelbracht {\it et al.} (1998)) is dominated by red supergiants (the
progenitors of core-collapse SNe).  We assume that this is roughly
appropriate for NGC~5962 which has an $L_{\rm FIR}$ higher than
NGC~253, and is classified as a ``nuclear HII-region galaxy'' (Misselt
et al. 1999) indicating a moderate level of nuclear star formation.
We therefore used the nuclear $K$-band light profile as
being roughly representative of the SN spatial distribution.  Random
distributions of ($x,y$) coordinates were generated for the artificial
supernovae, weighted according to the observed $K$-band light
distribution of NGC~5962 within 500~pc of the centre.

\subsection{Simulation of a Supernova Search}
We used the simulated occurrence of the supernovae together with the 
supernova detection efficiency estimated
above to predict the number of supernovae discoverable during an
observing programme using a $\sim$3m telescope (e.g. the NASA Infrared 
Telescope Facility or the Nordic Optical Telescope).  We considered a sample 
of 45 galaxies (see Table~A1) observable in the northern hemisphere. 
These 45 galaxies were selected according to their far-IR luminosities, 
far-IR colours and distances (see Appendix~A).

In Figure $\ref{fig7}$ we show, for a fixed integration time per
galaxy, the fraction of simulated SNe detected in a single observation
as a function of time interval, $t_{intv}$, since the previous
observation.  Each curve is based on the simulation of 100 000 SNe within the
innermost 500~pc radius of NGC~5962 (see Fig. 3 left) as described in
subsection~6.1.  The effects of intrinsic dispersion in the supernova
magnitudes and a 5$\%$ intrinsic fraction of slow-decliners have been
included.  Results for a range of extinctions and distances are
illustrated.  For example, at 30~Mpc, for a time
interval of 200~days between successive observations, we find that if
$A_{V}$ is increased from 18 to 36 the fraction of SNe recovered is
reduced by a factor of $\sim$4.  Likewise, for an extinction of $A_{V}$ = 27,
increasing the simulated distance to the host galaxy from 20~Mpc to 40~Mpc 
cuts the fraction detected by a factor of $\sim$2.5.  These results are now 
used along with the predicted intrinsic SN rate ({\it cf.} Section~2) to 
examine how the number of discoverable SNe in a galaxy sample is affected by 
choice of $t_{intv}$.

In Figure $\ref{fig8}$ we show the number of supernovae discoverable
from the sample of 45 galaxies for a given observation
of the entire galaxy sample (assuming that the galaxy comparison images
have been already acquired) as a function of the time interval,
$t_{intv}$, since the previous observation.  Again, each curve is based on
the simulation of 100 000 SNe.  These figures are used
below to estimate the best observing strategy.  The
number of SNe discoverable is plotted in Fig~7 (left) for three
different extinction values assuming a 5$\%$ fraction of
slow-decliners, and in Fig~7 (right) for the M~82-like extinction
distribution (see Table~6, col.~6).  In Fig~7 (right) we also illustrate the
effect of the expected higher rate of the slowly-declining events in
the nuclear starburst environment.  Such an enhanced rate could be due
to the higher densities caused by greater stellar mass-loss, stellar
winds and other, nearby SNRs in the nuclear regions (e.g. Cid
Fernandes 1997).  A recent study of the SNR flux variation in M~82 by
Kronberg {\it et al.} (2000) gives indirect evidence of enhanced
circumstellar densities.  They found that only $\sim$25$\%$ of the
sources varied measurably during a 11.8 year period, with the
remaining $\sim$75$\%$ having exceptionally stable integrated flux
densities.  We conclude that, for $t_{intv}\sim6$~months, if about
40$\%$ of the nuclear SNe were slowly-declining, we would expect a
$\sim35\%$ higher discovery rate than would be achieved for a
$\sim$5$\%$ SNe~IIn fraction.

With limiting magnitudes similar to those of the NGC~5962 images used
in this feasibility study, the 45 sample galaxies (and sky flats)
could be observed in $\sim$20 hours with a 3--4m telescope ($K$-band).
In a single observation of the sample, the expected number of
supernova discoveries is 1.8, 1.0, 0.5, for extinctions of $A_{V}$ =
18, 27 and 36 respectively.  This result corresponds to a time
interval exceeding $\sim$6~months since the previous observation of
the sample (see Figure $\ref{fig8}$(left)).  If we assume that the
distribution of extinctions towards the SNe is similar to that of
M~82, we obtain lower and upper limits of 1.0 and 1.3 respectively for
the expected number of SN detections, given slow-decliner percentage
fractions of 5$\%$ and 40$\%$ respectively (see Figure
$\ref{fig8}$(right)).

The adoption of the extinction distribution based on M~82, which is an
edge-on galaxy, our conservative approach to simulating supernovae in
galaxies closer than 34~Mpc (see Section~6), and the likely higher
fraction of the slowly-declining events in the nuclear starbursts,
taken together suggest that our supernova discovery rate estimates are
probably lower limits.

However, even if no supernovae were discovered in an observing
programme, useful upper limits could be derived for the nuclear
supernova rate, assuming particular extinction values.  For example,
in a negative search consisting of 3, 4, or 5 observations of the sample, an
upper limit for the supernova rate would be 3.1, 2.1, or 1.6 FIRSRU at 
90$\%$ confidence, assuming an M~82-like extinction (Section~3), and a 
slow-decliner percentage fraction of 5$\%$.  Alternatively lower limits could 
be derived for the average extinction towards the nuclear supernovae, 
assuming our estimated intrinsic nuclear supernova rate of 2.7 FIRSRU.  For 
negative surveys comprising 3, 4, and 6 observations of the sample, the 
extinction lower limits would be A$_{K}$ = 2, 3, and 4 respectively, with
at least 90$\%$ confidence

\subsection{Observing Strategy}
Given a fixed amount of telescope time, what distribution of the time
will maximise the number of detected supernovae?  Clearly, as we
increase the integration time per galaxy we shall a) increase the
volume of space within which SNe can be detected, and b) increase the
amount of time after explosion during which we can detect a given
SN. However, is it better to apply the same integration time to all
galaxies, or to increase the integration time as we go to more distant
galaxies? Keeping the integration time the same means that we can
detect the much rarer nearby SNe at much later points on their light
curves and so reduce the chance of missing these particularly
interesting events.  The disadvantage is that this procedure will
sometimes ``waste'' time detecting a nearby SN while it is still at an
early point on its light curve and therefore very bright.  On the
other hand, if we adjust the integration time to give a detection
threshold which is at the same epoch on the light curve regardless of
the galaxy distance then we can save time on the nearer events and use
the time saved to increase the volume of space explored.  The
disadvantage is that we are more likely to miss the fewer nearby
events.  It can be shown that the latter strategy will yield a
somewhat higher supernova detection rate.  However, this ignores the
fact that as we go to greater distances, the supernovae become
increasingly difficult to resolve against the host galaxy nucleus.
When we take that into account, the benefit of varying the integration
time with distance diminishes.  This fact, together with the desire to
avoid missing nearby SNe suggests that working with a fixed
integration time per galaxy is a reasonable strategy.

Turning now to the question of the number of times we should observe a
given galaxy.  For a given amount of telescope time and a fixed,
uniform integration time (${\it i.e.}$ magnitude threshold) per
galaxy, if we double the number of observations it simply halves the
number of galaxies we can observe.  In another approach which would allow a 
doubling of the number of times a given galaxy was observed, we could maintain 
the number of galaxies observed but reduce the integration time on each by
a factor of two (ignoring overheads). This would effectively reduce
the number of distant SNe that would be detected.  The volume of space
accessible above the threshold would be reduced by $2^{0.75}=1.68$
while doubling the number of observations.  This would yield a modest
net gain in the detection rate of $\times2/1.68=\times1.2$.  
So, given the overheads, there would be nothing gained in going for a larger 
number of shorter observations.  Thus, it appears that there would be little 
to be gained in dividing up the telescope time into more than a few 
observations per galaxy.  

Another issue concerns the interval between successive observations
of a given galaxy, $t_{intv}$.  Inspection of Figure $\ref{fig8}$ reveals 
that, at first, the SN detection rate increases quite rapidly with $t_{intv}$.
 This is due to the fact that, as $t_{intv}$ is increased more supernovae will
occur in the interval and yet we are still able to detect many of
those SNe which exploded soon after the previous observation.  If we
work with a fixed detection threshold, then the nearer, but rarer, SNe
can be detected at much later points on their light curves.  As a rule
of thumb, therefore, we should set $t_{intv}$ to equal the light curve
duration over which we can detect the nearest events.  Less time than
this will reduce the number of the nearest events we can detect.  More
time than this does not increase the number of detectable ``ordinary''
events, but simply increases the duration of the programme.  It is
true that as we continue to increase $t_{intv}$, more slow-decliners
will be found (see Figure~$\ref{fig8}$(right)).  However, the conservative 
assumption is that these
form a small minority of the sample.  For an ``ordinary'' template
$K$-band light curve with $M_K$ peak at --18.6 and assuming $A_K=3$,
our UFTI images suggest that the nearest likely supernova
($d$$\sim$10~Mpc) could be detected out to $\sim$140~days
post-explosion in an integration time of 5--15 minutes on a 3--4~m
telescope. Thus we are looking at $t_{intv}$ values of at least a few
months.  Taking into account also the slow-decliners we conclude that 
telescope time is used most efficiently if the time intervals between 
the observations of a given galaxy are {\it at least} $\sim$6 months.

\section{Conclusion} 
The feasibility of detecting supernovae in the nuclear regions of
nearby starburst galaxies has been investigated.  This has entailed the
examination of expected SN rates, extinction, and SN magnitudes in
such regions.  An intrinsic SN rate of 2.7 $\times$ 10$^{-12}$ $L_{\rm
FIR}$ / $L_{\odot}$~~yr$^{-1}$ was estimated.  Extinction values
ranging between $A_{K}$ = 2 and 4 were judged to be typical. A
particularly detailed examination of the M~82 extinction distribution
was carried out. A mean $A_{V}$ of 24 was obtained from hydrogen
column density measurements.  Template core-collapse SN light curves in
$JHKL$ were assembled. Two types were identified - ordinary and
slow-declining.  The slow-decliners were found to be significantly
more luminous than the ordinary core-collapse SNe, even at early
times.  Using the Optimal Image Subtraction method, it was found that
supernovae outside the innermost $\sim$1'' of a galaxy can be easily
detected in $K$-band images.  Imaging of a sample of 45 galaxies on a
6 month cycle, would probably reveal at least 1 SN per cycle.  However,
the actual number of detections could be significantly higher if
slow-decliners are more common in nuclear regions.  Near-infrared
observations of SNe in nuclear starbursts will provide us with a
valuable new probe of the conditions and star formation rates in the
obscured starburst regions of galaxies.
 
\section*{Acknowledgments}
We are grateful to C. Alard for providing us with the Optimal Image
Subtraction program and his generous help in its usage which made
this study possible.  We are also grateful to A.~Fassia, M.~Hernandez
and T.~Geballe for carrying out the NIR observations of NGC~5962. We
thank A. Efstathiou, A. Fassia, T. Hawarden, L. Lucy, N. Neininger,  
S. Ryder and K. Wills for helpful discussions. We also thank N. Neiniger and 
A. Wei\ss~for the $N$(H$_{2}$) values for M~82 prior to publication.  
We are grateful to the anonymous referee for a number of helpful suggestions. 
Financial support was provided to S.M. by Jenny ja Antti Wihurin rahasto, 
Vilho, Yrj\"o ja Kalle V\"ais\"al\"an rahasto, Osk. Huttusen S\"a\"ati\"o and 
the British Council.  This research has made use of the NASA/IPAC 
Extragalactic Database (NED) which is operated by the Jet Propulsion 
Laboratory, California Institute of Technology, under contract with the 
National Aeronautics and Space Administration.

\appendix

\section[]{The Galaxy Sample for the Discussed Observing Programme}

Rowan-Robinson \& Crawford (1989) considered
the far-IR spectra of galaxies as arising from a mixture of 1) a cool
disc component, 2) a warm starburst component, and 3) a hot Seyfert
component. They determined the relative proportions, $\alpha_{j}$, of
the spectrum attributable to each of the components by fitting models
to the far-IR spectra of the galaxies. They plotted IRAS colour-colour
diagrams for the flux density ratios: S(25)/S(12) $v.$S(60)/S(25) and
S(100)/S(60) $v.$ S(60)/S(25). In these plots, the starburst galaxies
occupy well-defined areas and thus the far-IR flux ratios can be used
as selection criteria for the galaxy sample.

We selected galaxies from Soifer {\it et al.} (1987, 1989) and
Rowan-Robinson \& Crawford (1989) whose far-IR luminosity is greater
or comparable to those of M~82 and NGC~253, excluding galaxies whose
far-IR luminosity is powered by a population of old stars or an AGN.

\noindent
We used the following criteria for selecting the sample :\\ 

\noindent
1a) for galaxies with $d$ $\leq$ 25Mpc : $L_{\rm FIR}$ $>$ 1.7 $\times$
10$^{10}$ L$_{\odot}$ \\

\noindent
1b) for galaxies with 25Mpc $<$ $d$ $\leq$ 35Mpc : $L_{\rm FIR}$ $>$ 2.5 $\times$
10$^{10}$ L$_{\odot}$ \\                                                  

\noindent                                
1c) for galaxies with 35Mpc $<$ $d$ $\leq$ 45Mpc : $L_{\rm FIR}$ $>$ 3.2 $\times$
10$^{10}$ L$_{\odot}$ \\                                                       

\noindent
2a) $S$(25)/S(12) $>$ 2.2, $S$(60)/S(25) $<$ 8.1, $S$(100)/S(60) $<$ 2.2 \\

\noindent
2b) from Rowan-Robinson \& Crawford (1989) the galaxies with
$\alpha_{2}$ $>$ 0.6 \\

\noindent
3) Seyfert 1 galaxies listed in Veron-Cetty and Veron
(1998) were excluded. \\

The flux ratios, FIR luminosities ($L$(8-1000$\mu m$) (Sanders
\& Mirabel 1996) and distances (Tully 1988) for the selected galaxies
are presented in Table $\ref{sample}$, cols. 2-6. In cols. 7-8, a
lower and upper limit for the intrinsic SN rate are presented
according to (a) equation (4) with $\epsilon=1$, and (b) equation
(5). In estimating the number of supernovae discoverable in an
observing programme we assumed that, for most of the 45 sample
galaxies, the fraction of their far-IR luminosity which is powered by
young massive stars is similar to that of NGC~253 and M~82.  However,
for the Seyfert 2 galaxies in the sample we reduced this fraction by a
factor of two, with the SN rates correspondingly reduced.

\begin{table*}
\centering 
\begin{minipage}{180mm} 
\caption{The galaxy sample.  The flux 
densities are taken from Soifer {\it et al.} (1989) and Rowan Robinson
\& Crawford (1989), and the distances mostly from Tully 1988. For
MGC-05-17-009, MGC-05-18-003, ESO~320-G30, ESO~402-G26, NGC~3690,
UGC~3630 the distances were obtained from the recession velocities
with respect to the Local Group (H$_{\rm 0}$ = 70 kms$^{-1}$/Mpc). For
NGC~253 and M~82 the distances were adopted from de Vaucouleurs (1978)
and Tammann \& Sandage (1968) respectively, and for NGC~4038/39 the same 
distance was assumed as in Neff $\&$ Ulvestad (2000).  The far-IR luminosities 
($L$(8-1000$\mu m$)) were calculated from the given flux densities and distances 
according to Sanders $\&$ Mirabel (1996). The far-IR luminosities are converted
to the intrinsic SN rates according to (a) Equ. 5 and (b) Equ. 1 in
Section 2.  The values given in column (a) were obtained assuming
$\epsilon=1$ and so are probably lower limits.}
\begin{tabular}{@{}llllclll@{}} \hline
 Galaxy & $\frac{S(25)}{S(12)}$ & $\frac{S(100)}{S(60)}$ &
 $\frac{S(60)}{S(25)}$ & Log($L_{\rm FIR}$) &\multicolumn{1}{c}{d} & \multicolumn{2}{c}{r$_{SN,intr}$} \\ 
& & & & [L$_{\odot}$] &\multicolumn{1}{c}{[Mpc]} & (a) & (b) \\ \hline 
NGC~253 & 3.77 & 2.00 & 6.76 & 10.28 & 2.5 & 0.02 & 0.05 \\
NGC~470 & 3.37 & 1.69 & 5.14 & 10.41 & 33 & 0.03 & 0.07 \\
NGC~520 & 3.34 & 1.48 & 10.38 & 10.89 & 30 & 0.09 & 0.21 \\
NGC~660 & 2.68 & 1.45 & 9.07 & 10.53 & 13 & 0.04 & 0.09 \\
NGC~1022 & 4.62 & 1.37 & 5.80 & 10.36 & 20 & 0.03 & 0.06 \\
NGC~1068 & 2.33 & 1.30 & 2.16 & 11.31 & 15 & 0.12 & 0.28 \\
NGC~1222 & 3.78 & 1.18 & 5.77 & 10.61 & 33 & 0.04 & 0.11 \\
NGC~1482 & 3.05 & 1.29 & 7.47 & 10.65 & 21 & 0.05 & 0.12 \\
UGC~3630 & 3.65 & 1.75 & 7.30 & 10.27 & 23 & 0.02 & 0.05 \\
MGC-05-17-009 & 3.45 & 1.40 & 6.44 & 10.81 & 39 & 0.07 & 0.17 \\
MGC-05-18-003 & 3.10 & 1.73 & 7.99 & 10.68 & 39 & 0.05 & 0.13 \\
NGC~2339 & 3.99 & 1.68 & 8.83 & 10.77 & 33 & 0.06 & 0.16 \\
NGC~2799 & 4.0 & 1.42 & 6.84 & 10.73 & 29 & 0.06 & 0.14 \\
NGC~2964 & 2.39 & 1.94 & 6.85 & 10.35 & 23 & 0.02 & 0.06 \\
M~82 & 3.99 & 1.03 & 4.60 & 10.62 & 3.25 & 0.05 & 0.11 \\
NGC~3094 & 3.44 & 1.31 & 3.94 & 10.66 & 34 & 0.05 & 0.12 \\
NGC~3256 & 4.80 & 1.28 & 6.08 & 11.57 & 37 & 0.40 & 1.00 \\
NGC~3310 & 3.24 & 1.40 & 6.46 & 10.62 & 20 & 0.05 & 0.11 \\
NGC~3395 & 3.30 & 1.62 & 7.58 & 10.43 & 29 & 0.03 & 0.07 \\
NGC~3471 & 3.46 & 1.50 & 7.07 & 10.49 & 35 & 0.03 & 0.08 \\
NGC~3504 & 3.73 & 1.57 & 5.39 & 10.75 & 28 & 0.06 & 0.15 \\
NGC~3597 & 3.32 & 1.28 & 6.63 & 10.87 & 44 & 0.08 & 0.20 \\
NGC~3690 & 6.19 & 1.01 & 5.04 & 11.84 & 45 & 0.75 & 1.9 \\
NGC~3885 & 3.81 & 1.37 & 6.49 & 10.45 & 30 & 0.03 & 0.08 \\
ESO~320-G30 & 4.05 & 1.28 & 14.68 & 11.17 & 42 & 0.16 & 0.40 \\
NGC~4038/39 & 2.66 & 1.69 & 7.40 & 10.83 & 21 & 0.07 & 0.18 \\
NGC~4102 & 4.01 & 1.50 & 7.13 & 10.68 & 18 & 0.05 & 0.12 \\
NGC~4194 & 5.44 & 1.02 & 5.61 & 11.09 & 42 & 0.14 & 0.31 \\
NGC~4536 & 2.44 & 1.56 & 7.35 & 10.24 & 14 & 0.02 & 0.05 \\
NGC~4691 & 3.72 & 1.50 & 4.65 & 10.44 & 24 & 0.03 & 0.07 \\
NGC~5073 & 5.41 & 1.48 & 5.80 & 10.61 & 39 & 0.05 & 0.11 \\
NGC~5188 & 3.71 & 1.54 & 7.97 & 10.90 & 35 & 0.09 & 0.21 \\
NGC~5218 & 3.03 & 1.98 & 7.67 & 10.60 & 41 & 0.04 & 0.11 \\
NGC~5430 & 2.95 & 1.99 & 5.64 & 10.82 & 42 & 0.07 & 0.18 \\
NGC~5597 & 3.25 & 1.91 & 4.92 & 10.72 & 41 & 0.06 & 0.14 \\
NGC~5915 & 2.87 & 1.41 & 7.84 & 10.60 & 36 & 0.04 & 0.11 \\
NGC~5929 & 3.77 & 1.50 & 5.64 & 10.68 & 41 & 0.03 & 0.06 \\
NGC~6000 & 4.02 & 1.61 & 7.35 & 11.04 & 32 & 0.12 & 0.30 \\
NGC~6764 & 3.55 & 1.79 & 4.77 & 10.51 & 38 & 0.04 & 0.09 \\
NGC~6835 & 3.00 & 1.48 & 7.19 & 10.33 & 25 & 0.02 & 0.06 \\
ESO~402-G26 & 2.44 & 1.59 & 8.71 & 10.49 & 39 & 0.03 & 0.08 \\
NGC~7479 & 2.80 & 1.60 & 3.92 & 10.84 & 35 & 0.04 & 0.09 \\
NGC~7552 & 4.02 & 1.37 & 6.04 & 11.06 & 23 & 0.12 & 0.31 \\
NGC~7582 & 4.69 & 1.50 & 7.48 & 10.73 & 20 & 0.03 & 0.07 \\
NGC~7714 & 6.00 & 1.11 & 3.73 & 10.72 & 40 & 0.06 & 0.14 \\
\label{sample}
\end{tabular}
\end{minipage}
\end{table*}

\section[]{Infrared photometry of core-collapse supernovae}

\begin{table*}
 \centering
 \begin{minipage}{190mm}
  \caption{Infrared photometry of core-collapse supernovae}
  \begin{tabular}{@{}lllllll@{}} \hline
Supernova & Date   & J     & H & K & L & Source \\ \hline
SN 1979C  & 43987.8 & 11.56(5)  & 11.32(5)  & 11.16(5) & 10.61(7) & Merrill private communication \\    
          & 43999.8 & 11.71(5)  & 11.35(5)  & 11.16(5) & 10.65(7) & Merrill private communication \\
	  & 44006.8 & 11.84(5)  & 11.87(5)  & 11.29(5) & -     & Merrill private communication \\
          & 44008.8 & 11.78(5)  & 11.46(5)  & 11.25(5) & 11.28(33) & Merrill private communication \\
          & 44042.8 & 12.43(5)  & 12.21(5)  & 11.91(6) & 11.22(20) & Merrill private communication \\
          & 44241.8 & 15.15(7)  & 13.66(6)  & 12.60(6) & 10.93(25) & Merrill private communication \\
          & 44262.8 & -        & -        & 12.75(7) & -     & Merrill private communication \\ \hline
SN 1980K  & 44544.5 & 10.46(5) & 10.23(5) & 10.07(10 & 9.66(10) & Dwek et al. (1983)\\
          & 44545.5 & 10.49(5) & 10.40(5) & 10.18(5) & 9.92(15) & Dwek et al (1983)\\
          & 44547.5 & 10.40(5) & 10.28(5) & 10.10(5) & 9.88(6) & Dwek {\it et al.} (1983)\\
          & 44551.5 & 10.58(5) & 10.43(5) & 10.24(5) & 9.97(5) & Dwek {\it et al.} (1983)\\
          & 44552.5 & 10.55(5) & 10.42(5) & 10.22(5) & 9.89(6) & Dwek {\it et al.} (1983)\\
          & 44557.5 & 10.71(5) & 10.56(5) & 10.35(5) & 10.05(7) & Dwek {\it et al.} (1983)\\
          & 44558.5 & 10.60(5) & 10.53(5) & 10.33(5) & 10.00(15) & Dwek {\it et al.} (1983)\\
          & 44559.5 & 10.78(5) & 10.64(5) & 10.40(5) & 10.14(9) & Dwek {\it et al.} (1983)\\
          & 44576.5 & 11.13(5) & 10.93(5) & 10.64(5) & 10.35(7) & Dwek {\it et al.} (1983)\\
          & 44592.5 & 11.48(5) & 11.30(5) & 10.97(5) & 10.55(6) & Dwek {\it et al.} (1983)\\
          & 44755.5 & 16.71(14) & 16.09(19) & 14.69(18) & 12.60(15) & Dwek {\it et al.} (1983)\\
          & 44774.5 & 16.71(14) & 16.22(13) & 14.95(13) & -     & Dwek {\it et al.} (1983)\\ \hline
SN 1982E  & 45158.5 & 17.39(10) & 15.47(5) & 13.81(3) & - & Graham (1985) \\
          & 45175.5 & -         & 16.0(40) & 14.3(15) & - & Graham (1985) \\
          & 45184.5 & 18.28(9)  & 15.96(4) & 14.19(3) & - & Graham (1985) \\
          & 45225.5 & -         & -        & 14.91(6) & - & Graham (1985) \\
          & 45248.5 & -         & 17.78(11) & 15.60(5) & - & Graham (1985) \\
          & 45304.5 & -         & -        & 17.1(30) & -  & Graham (1985) \\ \hline
SN 1982R  & 45269.6 & 14.0(10)  & 14.2(10) & 13.9(10) & -  & Muller (1982) \\
          & 45304.5 & 15.29(4)  & 15.06(4) & 14.71(6) & -  & Graham (1985) \\
          & 45329.5 & 16.04(5)  & 15.92(7) & 15.31(6) & 13.0(30) & Graham (1985) \\
          & 45330.5 & 15.87(5)  & 16.06(7) & 15.34(6) & -  & Graham (1985) \\ \hline
SN 1983I  & 45493.8 & 14.39(11) & 13.93(7) & 13.90(5) & - & Elias {\it et al.} (1985) \\
          & 45496.6 & 14.47(6) & 13.97(4) & 13.96(3) & - & Elias {\it et al.} (1985) \\
          & 45538.7 & 15.91(6) & 15.15(4) & 15.16(7) & - & Elias {\it et al.} (1985) \\
          & 45539.7 & 15.80(8) & 15.08(5) & 15.13(5) & - & Elias {\it et al.} (1985) \\
          & 45543.7 & 16.10(7) & 15.22(5) & 15.20(6) & - & Elias {\it et al.} (1985) \\ \hline
SN 1983N  & 45527.3 & 11.06(3) & 11.17(3) & 10.87(3) & - & Panagia {\it et al.} in preparation \\ 
          & 45528.3 & 10.96(3) & 11.16(3) & 10.87(5) & - & Panagia {\it et al.} in preparation \\
          & 45529.3 & 10.84(3) & 10.96(2) & 10.76(3) & - & Panagia {\it et al.} in preparation \\
          & 45531.8 & -        & -        & 10.64(3) & - & Panagia {\it et al.} in preparation \\
          & 45534.3 & 10.45(3) & 10.61(3) & 10.31(3) & - & Panagia {\it et al.} in preparation \\
          & 45538.8 & 10.42(3) & 10.38(3) & 10.36(3) & 10.17(5)$\footnote{The $L$-band magnitudes given for SN 1983N were observed in
$L'$-band.}$ & Panagia {\it et al.} in preparation \\
          & 45540.0 & 10.49(2) & 10.32(2) & 10.21(2) & 10.21(20) & Panagia {\it et al.} in preparation \\
          & 45540.7 & 10.55(6) & 10.37(5) & 10.23(5) & 10.20(20) & Panagia {\it et al.} in preparation \\
          & 45541.7 & 10.47(6) & 10.32(5) & 10.19(5) & 10.04(13) & Panagia {\it et al.} in preparation \\
          & 45559.5 & 11.05(6) & 10.76(7) & 10.91(13) & - & Panagia {\it et al.} in preparation \\
          & 45561.5 & 11.11(6) & 10.95(8) & 10.75(12) & - & Panagia {\it et al.} in preparation \\
          & 45577.5 & 11.66(10) & 11.31(14) & 11.41(23) & - & Panagia {\it et al.} in preparation \\
          & 45738.0 & -         & - 	  & 14.58(8) & - & Elias {\it et al.} (1985) \\
          & 45807.8 & - 	& - 	  & 15.78(14) & - & Elias {\it et al.} (1985) \\
          & 45829.8 & - 	& - 	  & 16.77(44) & - & Elias {\it et al.} (1985) \\
          & 45862.7 & - 	& - 	  & 17.52 (94) & - & Elias {\it et al.} (1985) \\ \hline
SN 1984L  & 45959.7 & 13.26(9) & 13.07(8) & 12.87(7) & - & Elias {\it et al.} (1985) \\
          & 45979.7 & 13.63(4) & 13.32(4) & 13.20(4) & - & Elias {\it et al.} (1985) \\
          & 46008.6 & 14.49(6) & 14.00(4) & -        & - & Elias {\it et al.} (1985) \\
          & 46009.6 & 14.42(4) & 13.91(4) & 13.97(4) & - & Elias {\it et al.} (1985) \\
          & 46050.6 & 15.62(8) & 14.88(5) & 15.05(4) & - & Elias {\it et al.} (1985) \\ \hline
SN 1985L  & 46265.5 & 12.62(2) & 12.47(2) & 12.38(3) & 12.22(41) & Meikle private communication\\
          & 46292.8 & 12.96(5) & 12.81(5) & 12.70(7) & 12.87(65) & Meikle private communication\\
          & 46293.8 & -        & -        & 12.75(7) & 12.81(63) & Meikle private communication\\
          & 46302.5 & 13.21(1) & 12.92(4) & 12.95(4) & 12.75(74) & Meikle private communication\\ \hline
\end{tabular}
\end{minipage}
\end{table*}

\begin{table*}
 \centering
 \begin{minipage}{190mm}
  \begin{tabular}{@{}lllllll@{}} \hline
Supernova & Date   & J     & H & K & L & Source \\ \hline
SN 1985P  & 46419.1 & 13.02(2) & 12.88(2) & 12.61(2) & -         & Meikle private communication\\
          & 46477.1 & 15.29(13) & 15.08(12) & 14.70(17) & -        & Meikle private communication\\
          & 46518.1 & 15.46(4)  & 15.49(9)  & 15.04(10) & -        & Meikle private communication\\
          & 46520.1 & 15.90(3)  & 15.92(7)  & 15.36(13) & -        & Meikle private communication\\
          & 46594.1 & 17.32(9)  & 17.74(27)  & 17.9(1.4) & -  & Meikle private communication\\ \hline
SN 1990E  & 47938.6 & 14.56(20) & 14.44(20) & - & - & Schmidt {\it et al.} (1993) \\
          & 47939.6 & 14.42(20) & 14.26(20) & - & - & Schmidt {\it et al.} (1993) \\
          & 47943.6 & 14.14(20) & 14.21(20) & - & - & Schmidt {\it et al.} (1993) \\
          & 47965.6 & 13.57(20) & 13.44(20) & - & - & Schmidt {\it et al.} (1993) \\ \hline
SN 1993J  & 49078.4 & 10.57(5) & 10.48(5) & 10.41(5) & - & Calamai {\it et al.} (1993) \\
          & 49078.7 & 10.54(2) & 10.47(2) & 10.37(2) & - & Odewahn {\it et al.} (1993)  \\
          & 49079.8 & 10.62(1) & 10.49(1) & 10.33(1) & - & Odewahn {\it et al.} (1993) \\
          & 49080.9 & 10.91(5) & 10.66(5) & 10.36(8) & - & Lawrence {\it et al.} (1993) \\
          & 49091.7 & 10.22(2) & 10.15(2) & 9.91 (2) & - & Romanishin {\it et al.} (1993) \\
	  & 49092.7 & 10.18(2) & 10.15(2) & 9.87(2) & - & Romanishin {\it et al.} (1993) \\
          & 49093.6 & 10.13(2) & 10.09(2) & 9.89(2) & - & Romanishin {\it et al.} (1993) \\
          & 49094.6 & 10.10(2) & 10.07(2) & 9.83(2) & - & Romanishin {\it et al.} (1993) \\
          & 49095.6 & 10.05(2) & 10.05(2) & 9.82(2) & - & Romanishin {\it et al.} (1993) \\
          & 49096.6 & 10.04(2) & 10.04(2) & 9.82(2) & - & Romanishin {\it et al.} (1993) \\
          & 49097.7 & 10.02(3) & 10.02(3) & 9.84(3) & - & Smith {\it et al.} (1993) \\
          & 49098.7 & 10.01(5) & 10.00(5) & 9.80(5) & - & Lawrence {\it et al.} (1993) \\
          & 49098.8 & 10.05(3) & 10.02(3) & 9.85(3) & - & Smith {\it et al.} (1993) \\
          & 49099.7 & 10.05(3) & 9.97(3) & 9.85(3) & - & Lawrence {\it et al.} (1993) \\
          & 49099.7 & 10.06(3) & 10.02(3) & 9.87(3) & - & Smith {\it et al.} (1993) \\
          & 49101.6 & 10.14(3) & 10.04(3) & 9.89(3) & - & Smith {\it et al.} (1993) \\
          & 49102.7 & 10.16(1) & 10.09(3) & 9.89(3) & - & Smith {\it et al.} (1993) \\
          & 49103.0 & 10.15(6) & -        & -       & - & Wada \& Ueno (1997) \\
          & 49104.1 & 10.23(16) & -       & -       & - & Wada \& Ueno (1997) \\
          & 49105.1 & 9.79(38) & -        & -       & - & Wada \& Ueno (1997) \\
          & 49107.3 & -        & -        & 10.01(4) & - & Kidger {\it et al.} (1993) \\
          & 49109.3 & -        & -        & 10.20(3) & - & Kidger {\it et al.} (1993) \\
          & 49113.3 & -        & 10.32(4) & 10.16(4) & - & Kidger {\it et al.} (1993) \\
	  & 49114.3 & 10.87(4)  & 10.53(4) & 10.35(4) & - & Kidger {\it et al.} (1993) \\
          & 49120.8 &  10.92(2) & 10.48(3) & 10.48(3) & - & Lawrence {\it et al.} (1993) \\
          & 49122.2 & 10.81(8) & -         & -       & - & Wada \& Ueno (1997) \\
          & 49123.1 & 10.85(13) & -        & -       & - & Wada \& Ueno (1997) \\
          & 49124.2 & 10.88(6) & -         & -       & - & Wada \& Ueno (1997) \\
          & 49129.1 & 10.98(4) & -         & -       & - & Wada \& Ueno (1997) \\
          & 43134.0 & 11.11(9) & -         & -       & - & Wada \& Ueno (1997) \\
          & 49135.0 & 11.11(8) & -         & -       & - & Wada \& Ueno (1997) \\
          & 49160.7 & 12.45(9) & 11.69(6) & 11.20(9) & - & Lawrence {\it et al.} (1993) \\
          & 49175.7 & 13.03(15) & 12.13(7) & 11.73(12) & - & Lawrence {\it et al.} (1993) \\ 
          & 49183.7 & 12.92(9) & 12.51(6) & 11.89(10) & - & Lawrence {\it et al.} (1993) \\ \hline
SN 1994I  & 49459.0 & -        & -        & 12.56(4)$\footnote{The $K$-band magnitude given for SN 1994I was observed in
$K'$-band.}$  & - & Grossan {\it {\it et al.}} (1999) \\ \hline
SN 1998S  & 50889.0 & 12.11(1) & 12.06(1) & 12.05(2) & -  & Fassia {\it et al.} (2000) \\
          & 50890.0 & 12.07(1) & 12.02(1) & 11.98(2) & - & Fassia {\it et al.} (2000) \\
          & 50890.9 & 12.06(1) & 12.01(1) & 11.98(2) & - & Fassia {\it et al.} (2000) \\
          & 50891.9 & 12.02(1) & 11.97(1) & 11.93(4) & - & Fassia {\it et al.} (2000) \\
          & 50892.9 & 12.02(1) & 11.93(2) & 11.90(2) & - & Fassia {\it et al.} (2000) \\
          & 50924.9 & 12.37(1) & -        & -        & - & Fassia {\it et al.} (2000) \\
          & 50951.9 & 13.36(2) & 13.05(3) & 12.70(4) & - & Fassia {\it et al.} (2000) \\
          & 50976.8 & 14.25(2) & 13.92(2) & 13.31(5) & - & Fassia {\it et al.} (2000) \\
          & 50984.8 & 14.33(1) & 14.02(2) & 13.43(11) & - & Fassia {\it et al.} (2000) \\
          & 51003.9 & 14.73(6) & 14.34(6) & 13.55(1) & - & Fassia {\it et al.} (2000) \\
          & 51004.9 & -        & -        & 13.55(1) & 11.1(15) & Fassia {\it et al.} (2000) \\ 
	  & 51200.9 & 17.4(25)  & 14.5(15)  & 12.89(10) & 11.7(25) & Fassia private communication \\
          & 51240.9 & 17.2(25)  & 15.2(15)  & 13.10(10) & 11.6(25) & Fassia private communication \\
          & 51272.2 & 17.17(25) & 15.3(15)  & 13.54(10) & 11.6(25) & Fassia private communication \\
          & 51331.2 & 17.75(25) & 15.56(15) & 13.80(10) & 11.6(25) & Fassia private communication \\ \hline
\end{tabular}
\end{minipage}
\end{table*}

\bsp

\label{lastpage}

\end{document}